\documentclass[12pt,preprint]{aastex}

\def \etal{{et~al.\null}}

\def\lea{\mathrel{<\kern-1.0em\lower0.9ex\hbox{$\sim$}}}
\def\gea{\mathrel{>\kern-1.0em\lower0.9ex\hbox{$\sim$}}}
\newcommand{\lta}{{\>\rlap{\raise2pt\hbox{$<$}}\lower3pt\hbox{$\sim$}\>}}
\newcommand{\gta}{{\>\rlap{\raise2pt\hbox{$>$}}\lower3pt\hbox{$\sim$}\>}}

\shorttitle{Luminsity, Mass, and Age Distributions of M83 Clusters}
\author{Rupali Chandar,\altaffilmark{1} 
Bradley C. Whitmore,\altaffilmark{2} 
Hwihyun Kim,\altaffilmark{3} 
Catherine Kaleida,\altaffilmark{3} 
Max~Mutchler,\altaffilmark{2} 
Daniela~Calzetti,\altaffilmark{4}  
Abhijit~Saha,\altaffilmark{5} 
Robert O'Connell,\altaffilmark{6} 
Bruce~Balick,\altaffilmark{7} 
Howard~Bond,\altaffilmark{2} 
Marcella~Carollo,\altaffilmark{8} 
Michael~Disney,\altaffilmark{9} 
Michael~A.~Dopita,\altaffilmark{10},
Jay~A.~Frogel,\altaffilmark{11} 
Donald~Hall,\altaffilmark{12} 
Jon~ A.~Holtzman,\altaffilmark{13} 
Randy~A.~Kimble,\altaffilmark{14} 
Patrick~McCarthy,\altaffilmark{15} 
Francesco~Paresce,\altaffilmark{16} 
Joe~Silk,\altaffilmark{17} 
John~Trauger,\altaffilmark{18} 
Alistair~R.~Walker,\altaffilmark{19} 
Rogier~A.~Windhorst,\altaffilmark{3} 
and Erick~Young\/\altaffilmark{20}}

\shortauthors{Chandar et al.}
\email{Rupali.Chandar@utoledo.edu}
\altaffiltext{1}{Department of Physics \& Astronomy, The University of Toledo, Toledo, OH 43606}
\altaffiltext{2}{Space Telescope Science Institute, Baltimore, MD, USA}
\altaffiltext{3}{School of Earth and Space Exploration, Arizona State University, Tempe, AZ 85287-1404, USA}
\altaffiltext{4}{Dept. of Astronomy, University of Massachusetts, Amherst, MA 01003, USA}
\altaffiltext{5}{NOAO, Tucson, AZ 85726-6732, USA}
\altaffiltext{6}{Dept. of Astronomy, University of Virginia,  Charlottesville, VA 22904-4325, USA}
\altaffiltext{7}{Dept. of Astronomy, University of Washington,  Seattle, WA 98195-1580, USA}
\altaffiltext{8}{Institute of Astronomy, ETH-Zurich, Zurich, 8093  Switzerland}
\altaffiltext{9}{Dept. of Physics and Astronomy, Cardiff University,  Cardiff CF24 3AA, United Kingdom}
\altaffiltext{10}{Research School of Astronomy \& Astrophysics, The  Australian National University, Cotter Road, Weston Creek, ACT  2611, Australia}
\altaffiltext{11}{AURA, Washington, DC 20005, USA}
\altaffiltext{12}{Institute for Astronomy, Honolulu, HI 96822, USA}
\altaffiltext{13}{New Mexico State University, Las Cruces, NM 88003,  USA}
\altaffiltext{14}{Goddard Space Flight Center, Greenbelt, MD 20771, USA}
\altaffiltext{15}{Carnegie Institute of Washington, Pasadena, CA  91101-1292, USA}
\altaffiltext{16}{ESTEC, Noordwijk, AG NL-2200, The Netherlands}
\altaffiltext{17}{Dept. of Physics, University of Oxford, Oxford  OX1 3PU, United Kingdom}
\altaffiltext{18}{NASA JPL, Pasadena, CA 91109, USA}
\altaffiltext{19}{Cerro Tololo Inter-American Observatory,
La Serena, Chile}
\altaffiltext{20}{NASA--Ames Research Center, Moffett Field, CA 94035}

\begin{document}


\title{The Luminosity, Mass, and Age Distributions of Compact Star Clusters in M83 Based on  
HST/WFC3 Observations}

\begin{abstract}

The newly installed Wide Field Camera 3 (WFC3) on the {\em Hubble
Space Telescope} has been used to obtain multi-band images of the
nearby spiral galaxy M83.  These new observations are the deepest and
highest resolution images ever taken of a grand-design spiral,
particularly in the near ultraviolet, and allow us to better
differentiate compact star clusters from individual stars and to
measure the luminosities of even faint clusters in the $U$ band.  We
find that the luminosity function for clusters outside of the very
crowded starburst nucleus can be approximated by a power law, $dN/dL
\propto L^{\alpha}$, with $\alpha = -2.04 \pm 0.08$, down to
$M_V\approx-5.5$.  We test the sensitivity of the luminosity function
to different selection techniques, filters, binning, and aperture
correction determinations, and find that none of these contribute
significantly to uncertainties in $\alpha$.  We estimate ages and
masses for the clusters by comparing their measured \textit{UBVI},H$\alpha$
colors with predictions from single stellar population models.  The
age distribution of the clusters can be approximated by a power-law,
$dN/d\tau \propto \tau^{\gamma}$, with $\gamma=-0.9 \pm 0.2$, for
$M\gea \mbox{few}\times10^3~M_{\odot}$ and $\tau \lea 4 \times
10^8$~yr.  This indicates that clusters are disrupted quickly, with
$\approx80$--90\% disrupted each decade in age over this time.  The
mass function of clusters over the same $M$-$\tau$ range is a power
law, $dN/dM \propto M^{\beta}$, with $\beta=-1.94\pm0.16$, and does
not have bends or show curvature at either high or low masses.
Therefore, we do not find evidence for a physical upper mass limit,
$M_C$, or for the earlier disruption of lower mass clusters when
compared with higher mass clusters, i.e., mass-dependent disruption.
We briefly discuss these implications for the formation and disruption
of the clusters.

\end{abstract}


\section{Introduction}

At a distance of 4.5 Mpc, M83, nicknamed the ``Southern Pinwheel,''  is the nearest massive grand-design spiral galaxy. It is a mildly barred galaxy, with a Hubble type SAB(s)c.   Less famous than its northern counterparts M51 (8.2 Mpc) and M101 (7.4 Mpc), M83 has been less studied in general. In this work, we use observations taken with the Wide-Field Camera 3 (WFC3) to study this galaxy in unprecedented detail. The combination of higher spatial resolution, extensive wavelength coverage,  and improved photometric accuracy, especially in the ultraviolet and near-IR,  make these the best observations of a grand-design spiral galaxy ever taken.

The WFC3 was designed to cover a wavelength range from $\sim2000$\AA~to $1.6\mu$, unprecedented for any instrument on-board the {\em Hubble Space Telescope} (\textit{HST}). Following in the legacy of WFPC1, WFPC2, and ACS, the WFC3 has $\approx$50 times the discovery efficiency (quantum efficiency $\times$ field of view) in the UV of either WFPC2 or ACS.  Similarly, WFC3 has $\approx$30 times the discovery efficiency in the near-IR of NICMOS.  The WFC3 Science Oversight Committee (SOC) has been granted $\approx$200 orbits of non-proprietary data, and roughly half of this allocation has been used to study star formation in different environments in the nearby universe (Early Release Science project \# 1; ERS1). Investigating star formation takes advantage of the sensitivity of WFC3 in both the UV and near-IR, since most of the light from young massive stars is produced in the UV, yet obscuration from dust can hamper
the study of the youngest star forming regions. Ten targets (NGC~3603, 30~Dor, M82, NGC~4214, NGC~5128, M83, NGC~2841, NGC~4150, NGC~4592, and M85) were selected as representatives of a wide range of environments and distances, ranging from 6~kpc for the massive, young star cluster NGC~3603 in our own Galaxy to 15.3 Mpc for the galaxy M85. More details about the ERS1 project, and its counterpart ERS2 (``Panchromatic WFC3 Survey of Galaxies at Intermediate Z''), can be found at:
http://www.stsci.edu/hst/proposing/docs/WFC3-ERS.
 
A complement of seven broad-band filters, ranging from the UV (F225W) to 
the near-IR (F160W), and seven narrow-band filters, ranging from F373N [OIII] to F164N [FeII], were used to image a field covering the starburst nucleus of M83, in order to take advantage of the panchromatic capabilities of WFC3. The broad-band filters provide estimates for the physical properties of the stars and star clusters, including luminosity, age, mass, and extinction. The narrow band filters provide estimates of physical properties in the interstellar medium (ISM), including shock diagnostics, ionization parameters, and abundances. The spatial resolution of WFC3 is slightly better than that of its predecessors (i.e., $0.04^{\prime\prime}$~pix$^{-1}$ compared to $0.05^{\prime\prime}$~pix$^{-1}$ in ACS; $0.13^{\prime \prime}$~pix$^{-1}$ in the near-IR compared to $0.20^{\prime\prime}$~pix$^{-1}$ for the NIC3 camera on NICMOS). A second, adjacent field just to the north has just been observed with the same set of filters. Dopita \etal\/ (2010) used the same dataset as used here to detect and study 60 supernovae remnants (SNRs) in M83. In future papers we will present the age
distribution for compact star clusters in M83 in more detail (Chandar \etal\/ in preparation), the distribution of cluster and association sizes (Kaleida \etal\/ in preparation), the infrared properties of the stars and clusters 
(Whitmore \etal\/ in preparation), the star formation history of M83 from individual (field) stars (Kim \etal\/ in preparation), and the population of HII regions, and the characteristics of dust attenuation and shock ionization in the ISM (Calzetti \etal\/ in preparation).

In this paper we determine the luminosity function of compact star clusters in M83. We provide a quantitative assessment of the impact that various corrections and different methods for selecting clusters and separating them from stars have on the results. We also present preliminary mass and age distributions for the clusters. We address the following specific questions in this paper:
(1)~Are the cluster luminosity and mass functions well described by power-laws?
(2)~Does the shape of the luminosity function depend on environment (e.g., nuclear vs.\ non-nuclear region) or on age (e.g., selected by color)?
(3)~To what degree is the shape of the luminosity function affected by the method used to select the clusters, by the wavelength of the filters, by binning or by aperture corrections?
(4)~What is the shape of the age distribution for clusters in M83?
(5)~Is there any evidence for a physical upper mass limit with which clusters
can form or for the faster disruption of lower mass clusters when compared with higher mass clusters (i.e., for bends in the mass function at the high or low mass end)?

The plan for the rest of this paper is as follows: Section~2 presents the observations and basic photometry; Section~3 describes the procedures used to select the star clusters, and compares results for automatically and manually selected catalogs; Section~4 presents the luminosity function for the clusters and assesses the impact of selection, binning, filter, and
aperture correction, and Section~5 presents preliminary mass and age distributions. Section~6 discusses the shape of these distributions in terms of the formation and disruption of the clusters. We summarize our main results in Section~7.

\section{Data and Observations}

Observations of a $3.6\times3.6$~kpc$^{2}$ portion of M83 were made
with the WFC3 on \textit{HST}, as part of the ERS1 program 11360 (PI: O'Connell) in August, 2009, in the following broad-band filters:
F225W (``$UV$''; 1800 sec), F336W (``$U$''; 1890 sec), 
F438W (``$B$''; 1180 sec), F555W (``$V$''; 1203 sec), 
F814W (``$I$''; 1203 sec), F110W (``$YJ$''; 1798 sec), 
and F160W (``$H$''; 1798 sec). Additional observations were made in
narrow-band filters covering the following lines: [OIII] (F373N; 2400 sec), H$\beta$ (F487N; 2700 sec), [OII] (F502N; 2484 sec), H$\alpha$ (F657N; 1484 sec), [SII] (F673N; 1850 sec), Paschen$\beta$ (F128N; 1198 sec) and [FeII] (F164N; 2398 sec). Three or four separate exposures were taken at different
dithered positions for each of the broad- and narrow-band filters. In addition to the longer exposures, short 10 sec exposures were taken in the F439W, F555W, and F814W filters to obtain photometry for objects that saturated in the long exposures. We focus here on the \textit{UBVI},H$\alpha$ observations. In this work, we assume a distance of $4.5\pm0.2$~Mpc to M83 (Thim et~al.\ 2003), corresponding to a distance modulus of $m-M = 28.28\pm0.1$. This yields a pixel scale of 0.876 pc~pix$^{-1}$.

\begin{figure}
\plotone{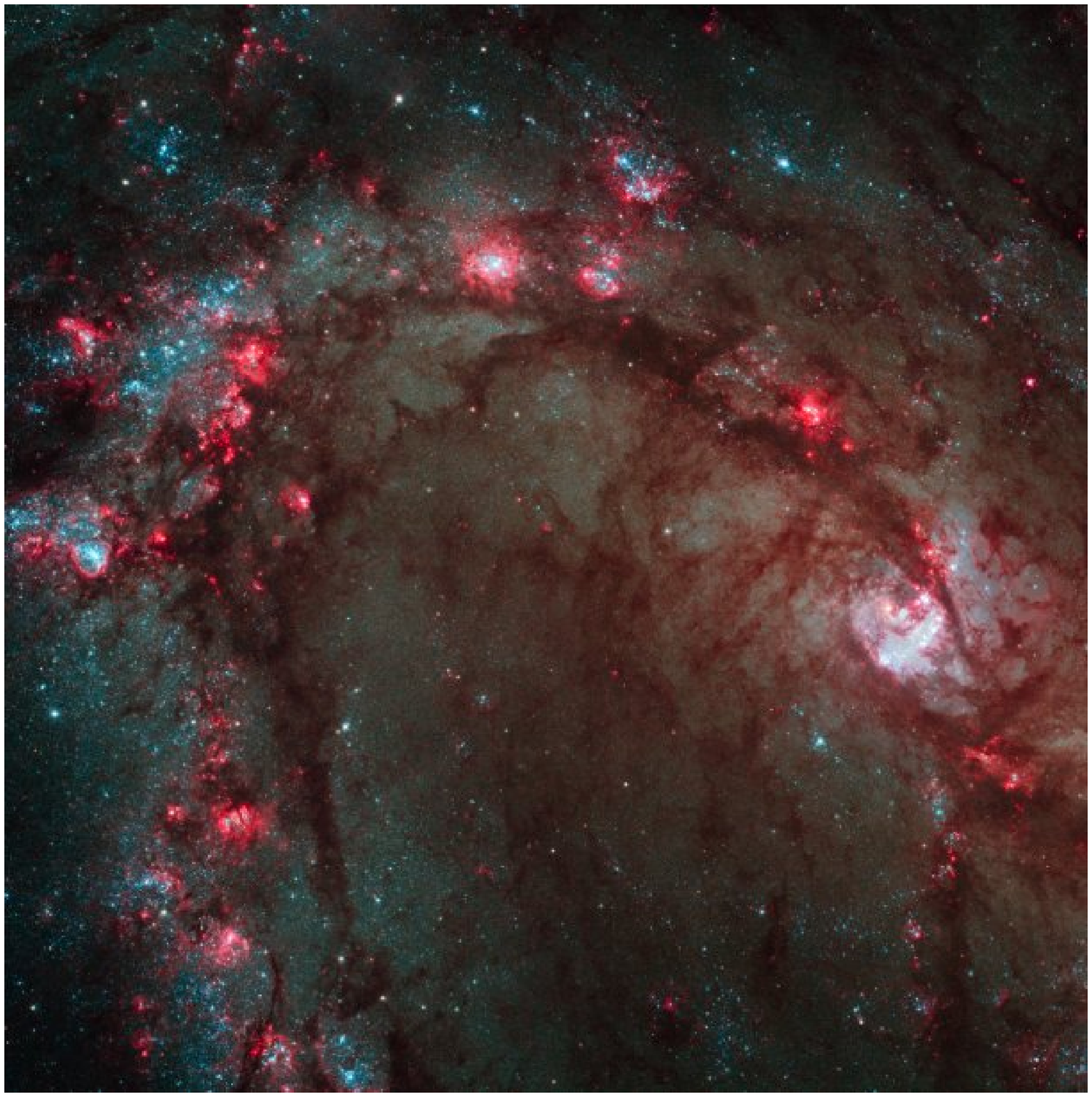}
\caption{Color image of M83 produced using the \emph{HST}/WFC3 observations described in this work.  The F438W image is shown in blue, the F555W image in green, and a combination of the F814W and H$\alpha$ images in red.}
\label{fig:colorimage}
\end{figure}

\begin{figure}
\plotone{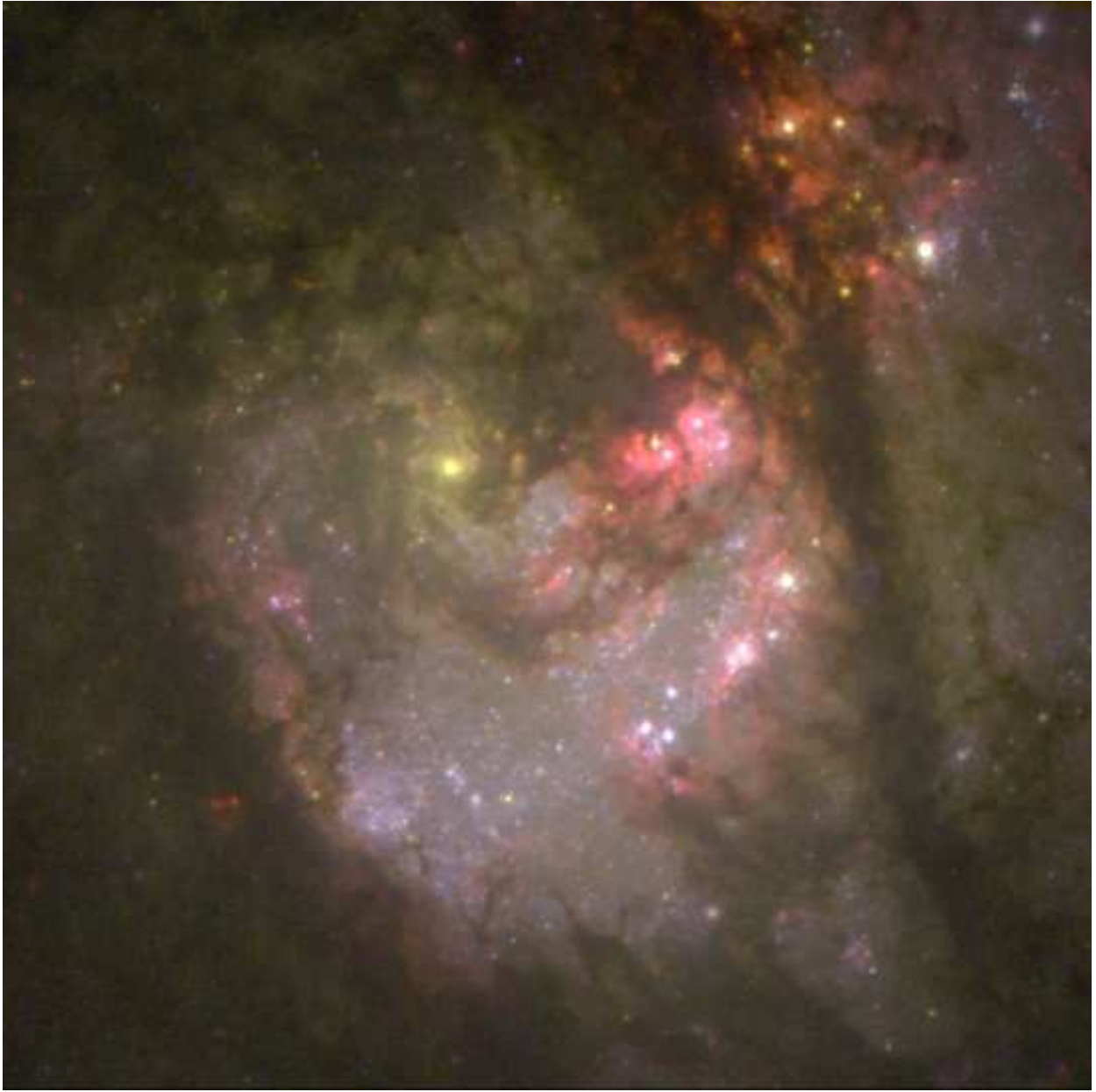}
\caption{Color image of the nuclear region of M83. }
\label{fig:nuclearimage}
\end{figure}

\begin{figure}
\plotone{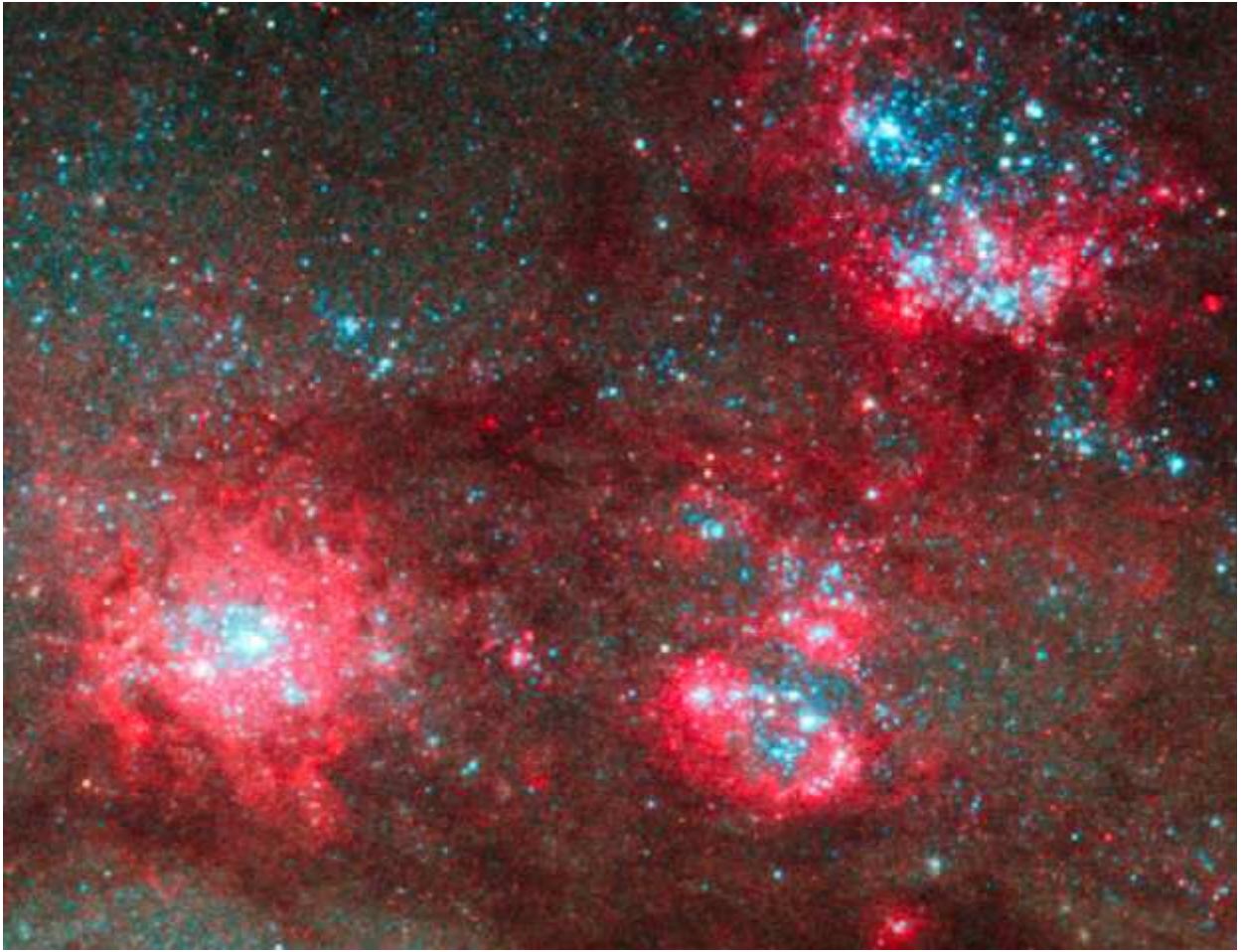}
\caption{Enlargement of a portion of M83 from Figure 1, showing a typical non-nuclear region including stars and star clusters.}
\label{fig:triangle_region}
\end{figure}

The flatfielded WFC3/UVIS images in each filter were co-added using the MULTIDRIZZLE task (Koekemoer et~al.\ 2002), with a final pixel scale of $0.0396^{\prime\prime}$~pix$^{-1}$ (based on the best calibration available for WFC3 at the time the data were reduced, and accurate to $\approx1$\%.)
A color image of the observed M83 field is shown in Figure~\ref{fig:colorimage}. The field covers the nuclear starburst region in M83, as well as a portion of a spiral arm and the inter-arm region. Figure~\ref{fig:nuclearimage} is an enlargement of the nuclear region, and shows that recent star formation has cleared out most of the dust over much of the area. 
A dust lane along the inner edge of the spiral arm is seen to the north and west of the nucleus, which appears as the  yellowish object in the upper left of the image. Figure~\ref{fig:triangle_region} shows an enlargement of a
typical portion of our field outside of the nuclear region, including three large star-forming complexes containing compact star clusters.

In order to detect as many sources as possible, we aligned and co-added together the final, drizzled broad-band images in the $U$, $B$, $V$, and $I$ filters, based on RMS-normalized images to give roughly equal weight to all wavelengths. This procedure allows us to include objects that are very blue or very red  in our source list, such as blue and red supergiant stars, that might otherwise be missing in any single filter. Using a median-divided  
version of the ``white light'' image (see discussion in Miller \etal\/ 1997), we identified all sources, both point-like and slightly extended, using the IRAF task DAOFIND. This resulted in a total list of $\approx 68,000$ objects, which includes a combination of individual stars, close blends of a few stars, star clusters, and background galaxies.

We perform circular aperture photometry of all detected sources on the drizzled images for each filter using the IRAF task PHOT, using an aperture radius of 3 pixels and a background annulus between 10 and 13 pixels. We have checked,  however, that our results for the luminosity and mass functions are not sensitive to the exact choice of aperture. For the narrow-band F657N ($H\alpha$) image, we perform photometry on the drizzled image without subtracting the stellar continuum flux. We convert the instrumental magnitudes to the VEGAMAG magnitude system by applying the following zeropoints:
F336W = 23.46, F438W =  24.98, F555W =  25.81,
F657N = 22.35, and F814W = 24.67, which are
provided by STScI at the following URL:
http://www.stsci.edu/hst/wfc3/phot\_zp\_lbn.
We will loosely refer to these as ``$UV$,'' ``$U$,''``$B$,'' ``$V$,'' and 
``$I$'' band magnitudes, although we do not make any transformations to the Johnson-Cousins system.

We use two different approaches to determine aperture corrections, which
convert our fixed aperture magnitudes to total magnitudes. Both approaches make use of the Concentration Index ($C$), measured in the $V$ band image and defined as the difference in aperture magnitudes determined using a 3 pix and a 0.5 pix radius. In the first approach, a single value (0.30) is used for the aperture correction of point sources (i.e., objects with  $C < 2.3$), and a different value (0.98) is used for extended sources (objects with $C > 2.3$) in each filter. These mean aperture corrections are determined from the magnitude difference between apertures of 3 and 12.5~pixel (i.e., $0.5^{\prime\prime}$) for a sample of $\approx$ 50 relatively
isolated, high S/N stars and clusters. We assume an additional 0.10 mag is needed to correct the photometry from $0.5^{\prime\prime}$ to infinity,
based on new measurements for WFC3. This assumption is good to a few percent and is dominated by the aperture correction, in any case. A weakness of this approach is that by adopting a single value of the aperture correction for all clusters, we overestimate the total luminosity of more compact clusters and underestimate the total luminosity of more
extended clusters.

Our second approach applies aperture corrections to each object based on its measured size ($C$). Unresolved objects (those with $C < 2.3$) all receive a single, filter-dependent aperture correction (0.30). For extended objects, we find the following relationship between aperture correction and $C$, for the range $ 2.3 < C \leq 3.4$: $m_\mathrm{apcorr}= -9.16 + 11.73\times C - 4.95\times C^2 + 0.71 \times C^3$. This relationship was determined by adding artificial star clusters of varying size, using the
MKSYNTH task in the BAOLAB package (Larsen 1999), to a relatively
uncrowded portion of the M83 image, and measuring their $C$ values and aperture corrections. A weakness of this approach is that values of $C$ can be quite uncertain for objects in crowded regions, where for example a compact cluster with a close neighbor can have an artificially large value of $C$, resulting in a too-high aperture correction, and leading to larger uncertainties in the luminosities determined for these clusters. Finally, we note that our color determinations use aperture magnitudes  rather than the more uncertain total magnitudes.

\section{Cluster Selection and Catalogs}

\subsection{Techniques}

\subsubsection{Using Size Information to Separate Stars and Clusters}

The selection of star clusters from observations that contain both individual stars and clusters is one of the most important steps for many studies of extragalactic cluster systems. In some cases the results of a study can be pre-determined, or at least strongly biased, by decisions made at this early 
phase of the analysis. One common situation is that samples of extragalactic clusters are far less complete at the youngest ages ($\tau \lea 10^7$~yr) than at older ages, because clusters tend to form in crowded regions where it is difficult to separate clusters from individual stars. These types of bias 
need to be accounted for since they can severely affect the luminosity, mass and age distributions of the clusters.

\begin{figure}
\epsscale{0.7}
\plotone{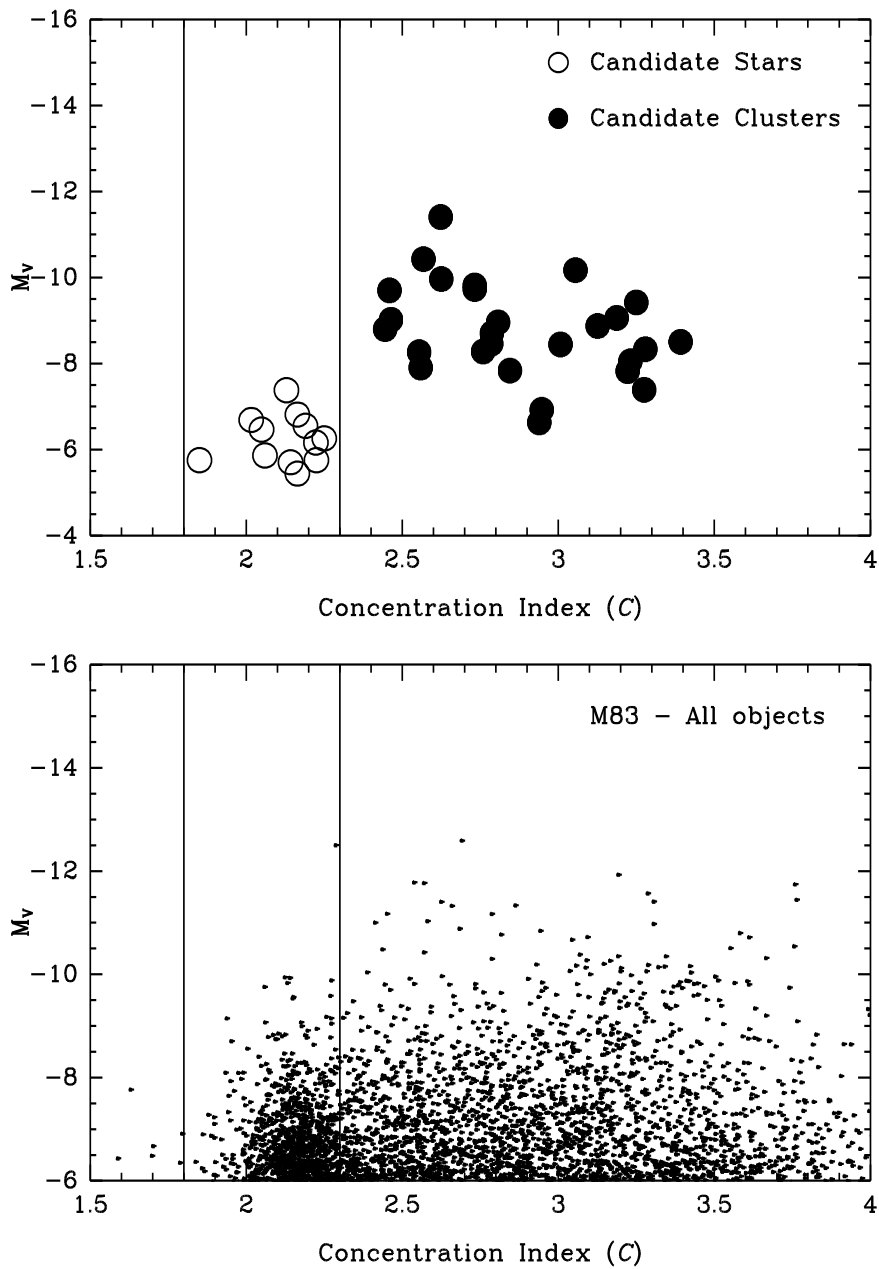}
\caption{Values of the concentration index $C$, the difference between magnitudes measured within a 0.5 and a three pixel radius, are plotted versus the absolute magnitude in the $V$ band (corrected for foreground extinction but not for extinction in M83).  The upper panel shows that hand-selected stars (open circles) and star clusters (filled circles) separate nicely, and the lower panel shows the distribution for all objects detected in our field. The vertical lines show the approximate range in $C$ found empirically for point sources. Note that the second brightest object in the lower panel, with $M_V\approx-12.7$, is a compact star cluster that falls just within the range of $C$ values used for stars.
}
\label{fig:trainingci}
\end{figure}

In our past work we have found that constructing ``training sets'' of stars and clusters is important for determining the best set of parameters to use 
for separating these objects (e.g., Whitmore \etal\/ 2010). Following this approach, we identified two training sets of objects in relatively isolated regions: a set of individual, point-like stars and a set of clearly extended, compact star clusters. The top panel in Figure~\ref{fig:trainingci} shows that
the concentration index $C$ cleanly separates these objects, with stars
having  $C < 2.2$, and clusters having $C > 2.3$. The bottom panel shows the corresponding figure for all objects brighter than $M_V \approx -6$. Here, the clean separation between stars and clusters evident for luminous isolated objects fills in at fainter magnitudes. We note that clusters dominate over stars in number at magnitudes brighter than $M_V \approx -8$, are comparable in number for $M_V \approx -7$, and stars dominate over clusters at magnitudes fainter than $M_V \approx -6$. We also note that most of the objects with $C > 3.5$ in Figure~\ref{fig:trainingci} are actually faint stars with bright nearby companions (see \S3.2.1), where the $C$ index is not able to measure the ``size'' accurately. Only 3\% of the objects  in Figure~\ref{fig:trainingci} which are brighter than $M_V \approx -5.5$ fall into this category. 

\begin{figure}
\epsscale{1.0}
\plotone{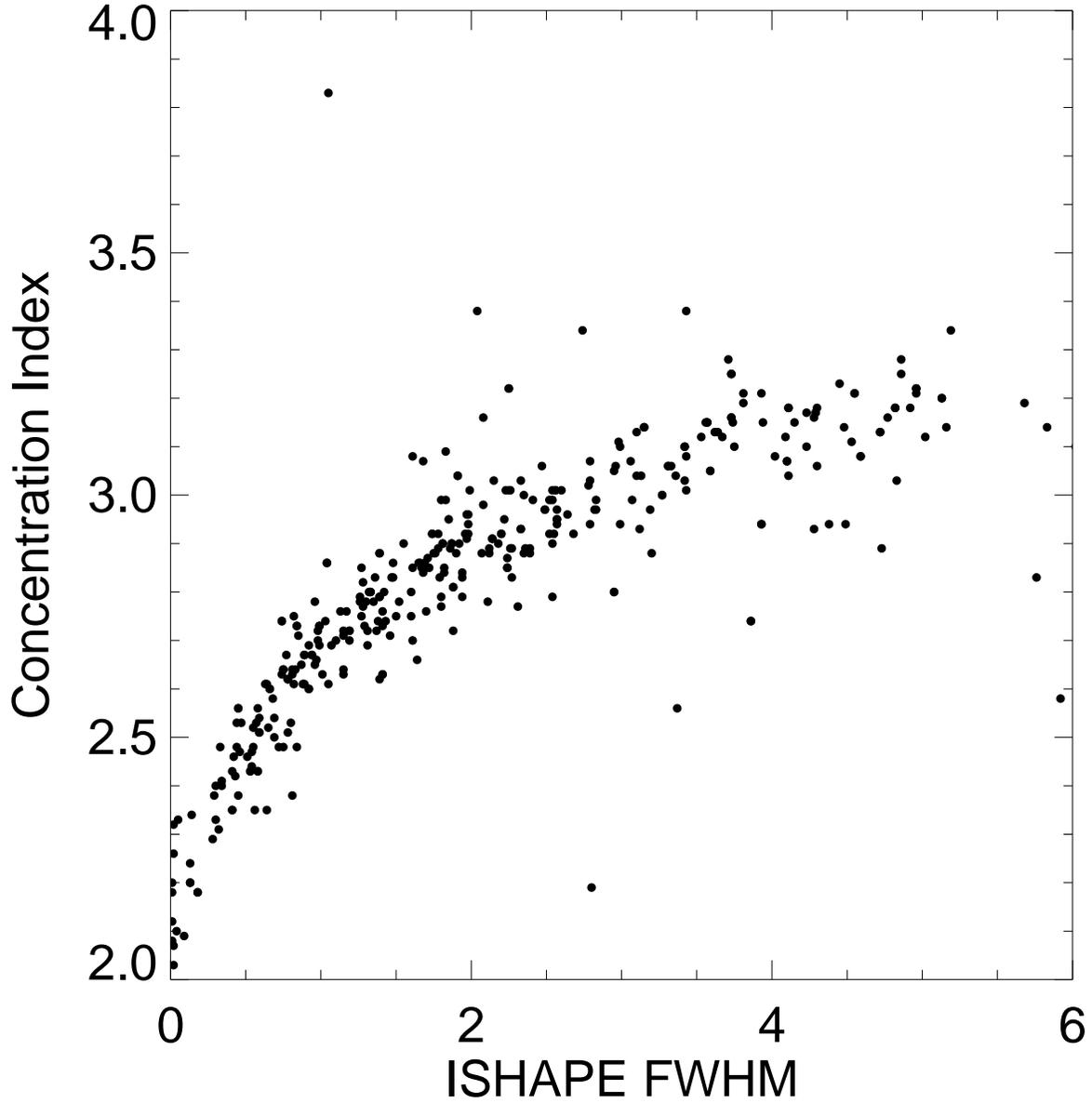}
\caption{Plot of two different size determinations for our hand-selected cluster catalog.  The concentration index $C$ is the same as measured in Figure~\ref{fig:triangle_region}. The FWHM (in pixels) determined by \emph{Ishape} is  based on the best fit to the two-dimensional image using an assumed King profile (1966) with a concentration of 30, convolved with an empirically determined PSF.}
\label{fig:ishape}
\end{figure}

In addition to $C$, we have used the \emph{Ishape} software, developed by S.~Larsen (Larsen 1999) to measure the sizes of objects. The $C$ index is a very simple method which works fairly well for most clusters (as discussed above). It is often more robust than more sophisticated methods, especially
in crowded regions and for faint objects. However, $C$  is also a fairly crude measure of object size, one which is less accurate than \emph{Ishape}
for isolated stars and clusters. \emph{Ishape} convolves analytic profiles with different values of the effective radius, $R_\mathrm{eff}$ (representing the surface brightness distribution of a star cluster) with the PSF, and determines the best fit to each source. We create a PSF in the F555W image from $\approx25$ relatively bright, isolated point sources, and use \emph{Ishape} to measure the $R_\mathrm{eff}$ of each source assuming a King profile (King 1966), with a ratio of tidal to core radius of 30, based on the flux within a 5 pixel radius. Figure~\ref{fig:ishape} shows a reasonably good correlation between $C$ and \emph{Ishape} measurements for the sizes of clusters in our manually selected catalog (described in Section~3.2.2), although the $C$ values ``saturate,'' i.e., stop increasing, before $R_\mathrm{eff}$ measured by \emph{Ishape} does. The size distribution of the clusters, along with correlations between the sizes, ages, and masses of clusters in M83, will be discussed in a future paper.

\subsubsection{Using Color Information to Separate Stars and Clusters}

\begin{figure}
\epsscale{0.75}
\plotone{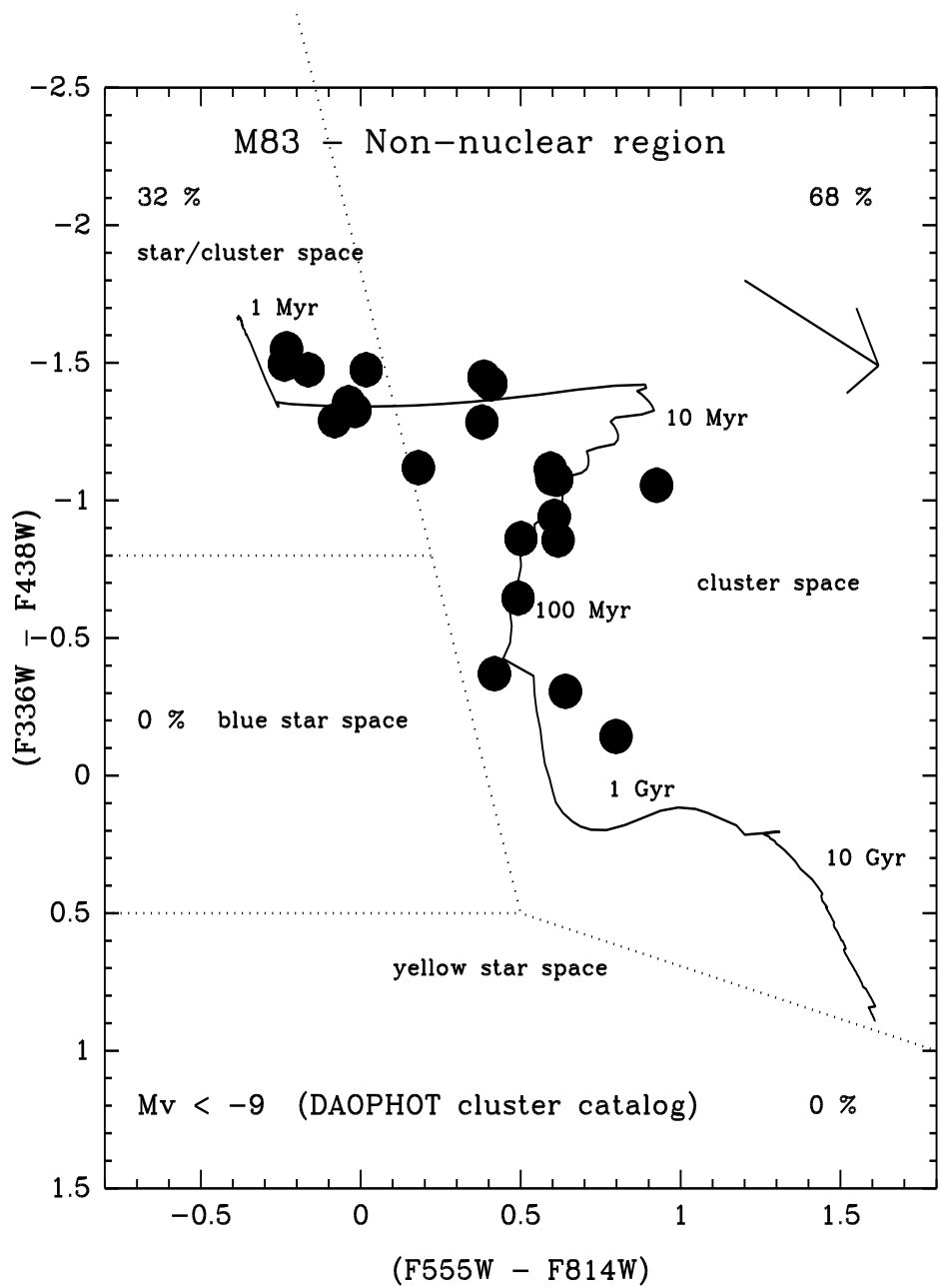}
\caption{($U\!-\!B$) vs.\ $(V\!-\!I)$ two-color diagram for cluster candidates 
outside of the nuclear region based on our Daophot catalog. The curves show predictions in the appropriate WFC3 filters from the single stellar
population models of Charlot \& Bruzual (2009) for twice solar (solid line) metallicity. Ages of $10^6$, $10^7$, $10^8$, $10^9$, and $10^{10}$ yr are
indicated. An  $A_V = 1.0$ reddening vector is shown in the upper right. See text for a discussion of how the diagram has been divided into cluster, star/cluster, blue star, and yellow star space. 
}
\label{fig:single_color}
\end{figure}

Optical colors, in addition to object size, can help to distinguish clusters from stars, as demonstrated in the Antennae galaxies by Whitmore \etal\/ (2010). 
In Figure~\ref{fig:single_color} we first compare $U\!-\!B$ vs.\ $V\!-\!I$ colors for clusters brighter than $M_V \leq -9$ from our Daophot catalog (described in Section~3.2.1), with new single stellar population model predictions from Charlot \& Bruzual (2009; hereafter CB09, private communication; also see Bruzual \& Charlot 2003) which use the WFC3 filter
transmission curves. The colors and luminosities have been corrected for
reddening and extinction in the Milky Way ($E(B-V)=0.07$, $A_V=0.218$; Schlegel \etal\/ 1998) but not in M83. We show predictions from models with 
$\approx2\times$solar metallicity (the inner portions of M83 have super-solar abundance; Bresolin et~al.\ 2005). We note the overall exquisite match between the observations and the models, greatly surpassing those of past observations in the $U$ band using WFPC2. This bodes well for our ability to age-date the clusters, which is discussed in Section~5. 

\begin{figure}
\epsscale{0.75}
\plotone{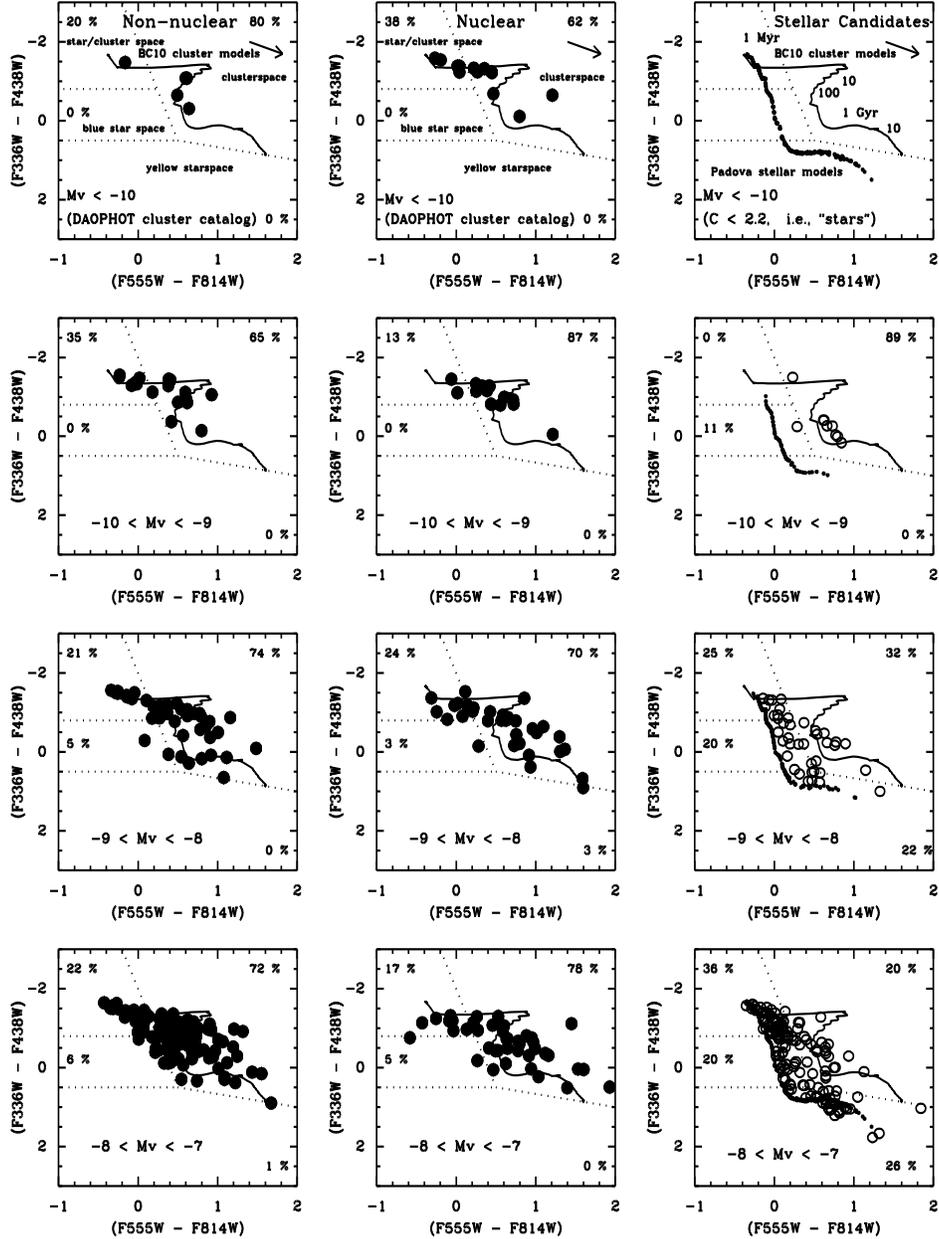}
\caption{($U\!-\!B$) vs.\ $(V\!-\!I)$ two-color diagrams for cluster candidates from our Daophot catalog from outside the nuclear region (first column of panels), for cluster candidates in the nuclear region (middle column of panels), and for stellar candidates with  $C < 2.2$  (right column of panels).  Each row shows objects in the indicated magnitude range,
starting with bright source at the top and moving to fainter objects at the bottom. The solid line shows predictions in the appropriate WFC3 filters from the single stellar population models of Charlot \& Bruzual (2009)  for two times solar metallicity. The small circles in the panels on the right show the predicted Padova isochrones with $\approx1.5\times$solar metallicity (the best match available to the cluster models) for individual stars of the indicated
luminosity.  The top panel however, shows predictions for all stars brighter than $M_V=-7$.
}
\label{fig:stvscl1}
\end{figure}

Next, we compare predicted colors for luminous stars from the Padova 
group\footnote{http://stev.oapd.inaf.it/cgi-bin/cmd} (e.g., Girardi et~al.\ 2002; Marigo et~al.\ 2008), shown as the small filled circles in the upper-right
panel of Figure~\ref{fig:stvscl1}, with the predicted cluster colors. Predicted colors for the stars begin at roughly the same location as for the youngest clusters, in the upper left portion of the two-color diagram, but the two tracks separate for clusters older than $\approx3\times10^6$~yr, when red supergiants begin to appear. We use the models as a guide to define four different regions of two-color space:

\noindent (1) ``cluster space'':  region of two-color space unique to clusters
($5\times(V\!-\!I) - (U\!-\!B) > 2$  and  $0.33\times (V\!-\!I) - (U\!-\!B) > -0.33$)

\noindent (2) ``star/cluster space'':  region of two-color space containing both very young clusters and very blue stars ($5\times(V\!-\!I) - (U\!-\!B) < 2$ and $(U\!-\!B) < -0.80$)

\noindent (3) ``blue-star space'':  region of two-color space unique to luminous, blue stars ($-0.80 < (U\!-\!B) < 0.50$ and $0.33\times (V\!-\!I) - (U\!-\!B) < -0.33$)

\noindent (4) ``yellow-star space'': region of two-color space unique to luminous, yellow stars ($(U\!-\!B) > 0.50$ and $0.33 \times (V\!-\!I) - (U\!-\!B) < -0.33$)

\noindent These four regions are somewhat different than those used in the Antennae by Whitmore \etal\/ (2010), primarily because of the significantly higher quality of the $U$ band observations provided by WFC3 compared with WFPC2.

Figure~\ref{fig:stvscl1} shows a series of two-color diagrams, divided into four different intervals of luminosity (down to $M_V = -7$).  The first two columns show our best (``Daophot''---see Section~3.2) catalog of star clusters in the non-nuclear (left) and nuclear (middle) regions, while the last column shows candidate stars, selected to have $C < 2.2$.  Note how most of the stars in the bottom right panels of Figure~\ref{fig:stvscl1} follow each bend in the Padova models almost exactly, with modest amounts of reddening moving some of the points slightly to the right of the models.  
There are however, a number of luminous stars (e.g., see the second panel down on the right) that are well to the right of the Padova models. We believe that these may be relatively rare, luminous blue variable stars (LBVs) in M83, since they have colors and luminosities similar to LBVs in the Galaxy and Magellanic Clouds.  For example, three of the best known LBVs, Eta
Carina, P~Cygni, and S~Doradus, have median colors of $U\!-\!B \approx-0.6$ and $V\!-\!I\approx0.8$ and $M_V$ between $-8.9$ and $-10.5$~mag.  The properties of luminous stars in M83 will be studied in more detail in Kim et~al.\ (in preparation).

A comparison between the two-color diagrams for clusters in the non-nuclear and nuclear regions (first and second column in Figure~\ref{fig:stvscl1}, respectively), shows strong similarities. In particular, the fraction of objects in each of the four regions of the two-color diagram, indicated in the figure, are within $\approx$10\% of each other.  This implies that the color distribution of the clusters is similar in the two regions. One difference is the higher extinction for many of the clusters in the nuclear region, which lie primarily along the reddening vector rather than on the model curves themselves.  Another difference is the larger scatter relative to the models and the smaller number of faint clusters in the nuclear region, as expected due to the higher background and hence lower completeness of the sample.

We can use Figure~\ref{fig:stvscl1} to assess how well our size-based ($C$) selection of clusters works. We find that 60--80\% of the resolved objects (candidate clusters) fall in cluster-space, while 20--40\% fall in star/cluster
space, independent of magnitude. Out of 189 candidate clusters selected
based on size, only  1 (1\%) fall (just barely) in yellow star-space, and only 9 (5\%) falls in blue star-space.  The primary remaining uncertainty is the fraction of misclassified objects in star/cluster space, either objects with measured $C>2.4$ which are blends of stars, or objects with $C < 2.2$ which are very compact star clusters. We address these issues in Section~3.2, and find there that any contamination in star/cluster space is at the 
$\approx10$\% level. We conclude therefore, that size measurements provide a relatively robust means of separating compact star clusters from individual stars in this dataset. This result is somewhat different from what we found in the more distant ($\approx20$~Mpc) Antennae galaxies, where using color as an additional critierion was more important for studying clusters fainter than $M_V\approx-8$ than it is here (see Whitmore \etal\/ 2010).

\subsection{Results}

Below, we build catalogs of star clusters selected in two different ways, automatically (using parameters measured from Daophot, \emph{Ishape},
and Sextractor) and manually.  The manual catalog provides an important consistency check on the automatic catalogs, and helps to quantify
contamination from blends, individual stars in crowded regions, and
background galaxies (the latter are negligible here).  We determine
systematic uncertainties in cluster selection by comparing the results from these catalogs. 

\subsubsection{Using Daophot and the Concentration Index to Produce a Cluster Catalog}

As outlined in \S3.1.1, our primary discriminant for separating individual stars and clusters is the Concentration index ($C$). A close inspection of candidate clusters selected based on size ($C > 2.4$) alone from our Daophot generated source list shows that there are a number of remaining contaminants, mostly due to close pairs of individual stars, in the most crowded regions. In many cases the overlapping profiles appear to cause the $C$ values to be artificially high, resulting in a large number of ``apparent'' clusters in a small area around the more populated clusters (e.g., the bottom left of Figure~\ref{fig:triangle_region}).  The colors of many of these sources, however, indicates that they are individual stars. We remove most of these cases by selecting the brightest object with $C > 2.4$  in the region and then removing any similar sources within 20 pixels. While this occasionally removes another cluster nearby, the vast majority of the removed objects are clearly close pairs of stars.  We can estimate the number of real clusters removed by this procedure by reinserting objects with $M_V < -10$, i.e. definite star clusters, since essentially all individual stars in nearby galaxies are fainter than this value (e.g., Humphreys \& Davidson 1979). We find that no clusters were removed by this procedure
in the non-nuclear region, while in the nuclear region approximately 20 clusters were removed due to the extreme crowding. These luminous clusters in the nuclear region are included in the catalog. The last remaining contaminant, even in relatively isolated regions, is from close pairs of stars of unequal brightness, where the fainter of the pair is measured to have $C>2.4$. We reject most of these remaining contaminants by removing  sources having brighter companions within 8 pixels.

The final catalog, which we refer to as our ``Daophot'' catalog, contains a total of 1247 cluster candidates brighter than $m_V\approx23.2$ (i.e., $M_V=-5.2$). Approximately 200 of these cluster candidates, mostly brighter than $m_V\approx22$, are in the crowded nuclear region, while the remaining candidates are found outside of the nuclear region. We consider the Daophot catalog to be our best cluster catalog, because a visual inspection shows that, overall, it is missing the fewest number of obvious clusters in both the nuclear and non-nuclear regions, and has the fewest number of false clusters such as from close pairs of stars.

Based on artificial cluster experiments, we find that the completeness of clusters in the nuclear region drops quickly for magnitudes fainter than
$m_V\approx 21.25$, more than a magnitude brighter than for the non-nuclear region. This is corroborated by the fact that the luminosity distributions for the clusters begin to flatten below these magnitudes.

\subsubsection{Using Ishape to Produce a  Cluster Catalog}

Size measurements can also be made using the \emph{Ishape} software (Larsen 1999), as described in Section~3.1.1. We measured the size of all $\approx68,000$ sources in our Daophot catalog using \emph{Ishape},  and selected cluster candidates to have \textit{FWHM} measurements between 0.5 and 10 pixels (i.e., at least 0.5~pixels broader than the point spread function) and a S/N of at least 50. We then applied the same neighbor criteria as used for the $C$-selected cluster catalog, to eliminate remaining contaminants. We find overlap between the $C$ and \emph{Ishape} selected
catalogs to be $\approx80$\%, but that \emph{Ishape} can fail to fit good clusters in more crowded regions, such as within star forming complexes
along the spiral arms and in the nuclear region.

The final catalog prepared by this approach, which we refer to as our \emph{Ishape} catalog, contains a total of 1130 cluster candidates brighter than $M_V\approx 23.2$. Approximately 180 of these candidates, mostly brighter than $m_V\approx 21.5$, are in the crowded nuclear region.

\subsubsection{Using SExtractor to Produce a Cluster Catalog}

\begin{figure}
\epsscale{1.0}
\plotone{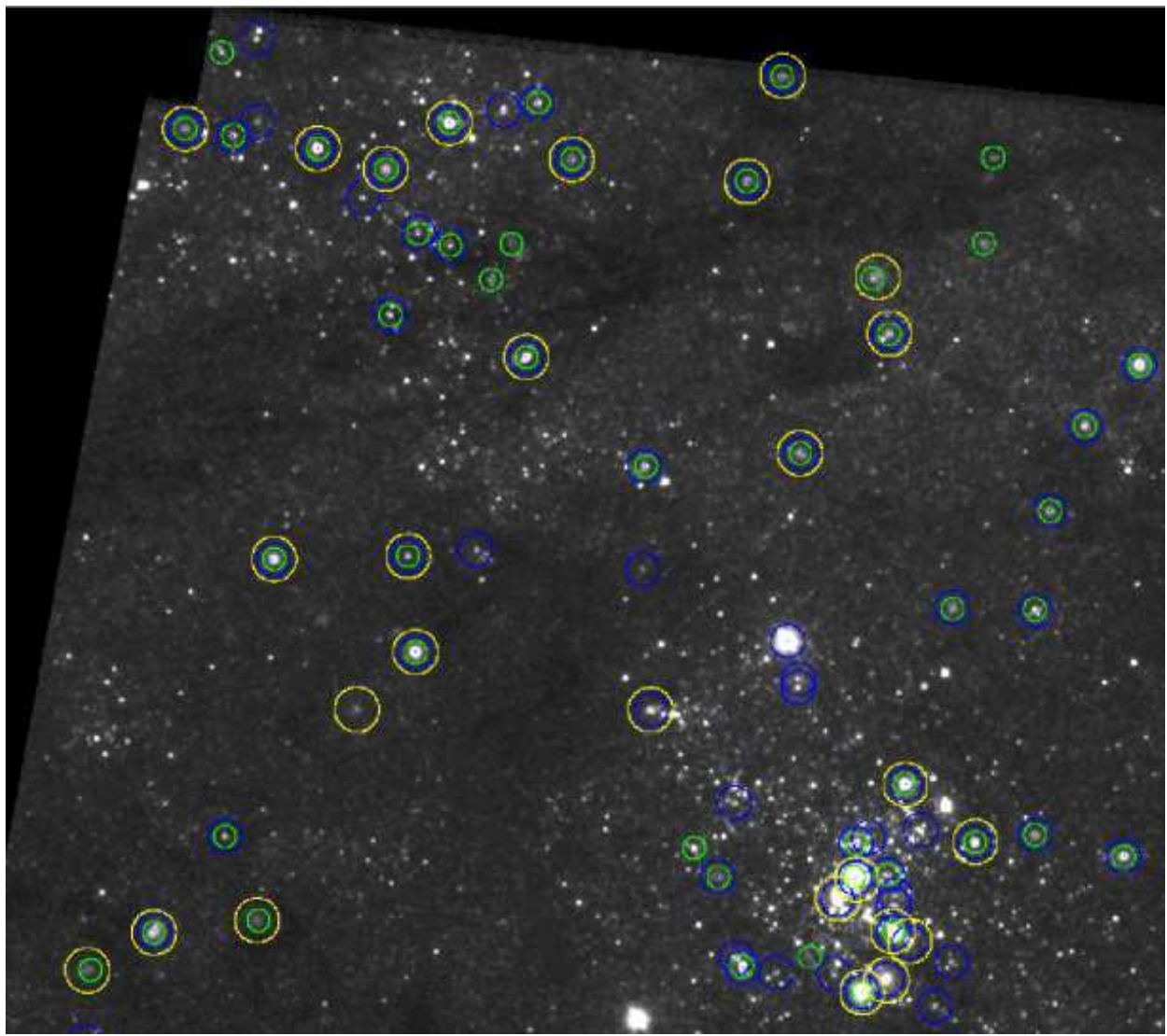}
\caption{A comparison of the Daophot (green), Sextractor (blue) and manual 
(yellow) cluster catalogs in a region of M83. The samples overlap at approximately the 80\% level, although the automatically selected catalogs extend to fainter sources.}
\label{fig:man_auto}
\end{figure}

The SExtractor software (Bertins \& Arnouts 1996) is also sometimes used to select star cluster candidates. We produced an independent source list using SExtractor, and selected  cluster candidates to have $\textit{FWHM} > 2$~pix (this \textit{FWHM} is not deconvolved from the PSF, as is done with \emph{Ishape}). Figure~\ref{fig:man_auto} shows a comparison between the SExtractor- and Daophot-based cluster catalogs (and the manual catalog that will be discussed in Section~3.2.4) for a small, relatively uncrowded region of our field.  We find that 36 of the 47 objects in our SExtractor catalog match objects in the Daophot catalog in this region (i.e., 77 \%).
SExtractor does a better job than Daophot of identifying star clusters as single objects rather than as multiple point sources, but tends to miss some of the more diffuse objects (e.g., two objects in the bottom left of Figure~\ref{fig:man_auto}), and also misses most of the very bright clusters in regions with high background, particularly in the nuclear region. We find that $\approx75$\% of all 901 objects brighter than $m_V=23$ and outside of the nuclear region in our SExtractor catalog match those in the Daophot catalog.

The final catalog prepared by this approach, which we refer to as our ``Sextractor'' catalog, contains a total of 1198 cluster candidates brighter than $m_V\approx23.2$. Approximately 200 of these cluster candidates,
mostly brighter than $m_V\approx22$ but extending to fainter magnitudes
than in the Daophot catalog, are in the crowded nuclear region.

\subsubsection{Manually Selecting a Cluster Catalog}

In addition to the various automatically selected catalogs, we also construct a catalog of clusters manually, by carefully examining the WFC3 images.  
This allows more difficult cases, such as a compact cluster near a bright star, to be assessed individually, which is not possible with automatically
selected samples.  We also found that this approach has the advantage of minimizing contamination by individual stars in crowded regions, at the expense of missing actual clusters in these regions. The primary shortcoming to this approach is that it is not possible to automatically reproduce the cluster
selection or to quantitatively determine the completeness of the sample.  Three of us (RC, HK, CK) selected clusters independently across the entire image, using slightly different approachs, but not pushing as deep as the automatic catalogs.  Our final, manually selected catalog contains 489 star clusters, mostly brighter than $m_V \approx 22.5$.

Figure~\ref{fig:man_auto} shows relatively good agreement between the manual and both the Daophot (21/27) and Sextractor (22/27) catalogs (i.e., $\approx$80\%) for clusters brighter than $m_V \lea 22.25$ ($M_V \lea -6$) as mentioned in the previous section, similar to the $\approx75$\% agreement found between the full catalogs outside of the nuclear region. The automatic catalogs are deeper, however, and contain more than double
the number of cluster candidates than the manual catalog. A careful inspection of the sources that are discrepant between the Daophot and manual catalogs ($\approx20$\% of the total) indicates that approximately half of these appear to be good clusters in crowded regions, while the other half are mainly remaining blends or superpositions of two close stars. This suggests that our automatic catalogs have contamination at the $\approx10$\% level.

\section{Luminosity Functions}

\subsection{Comparison Between Results from Different Catalogs}

In Section~3 we used four distinct approaches to select compact star clusters in M83. Here, we study the luminosity function of the clusters, and use the different catalogs to quantify the impact that different selection methods have on the results.

\begin{figure}
\epsscale{1.}
\plotone{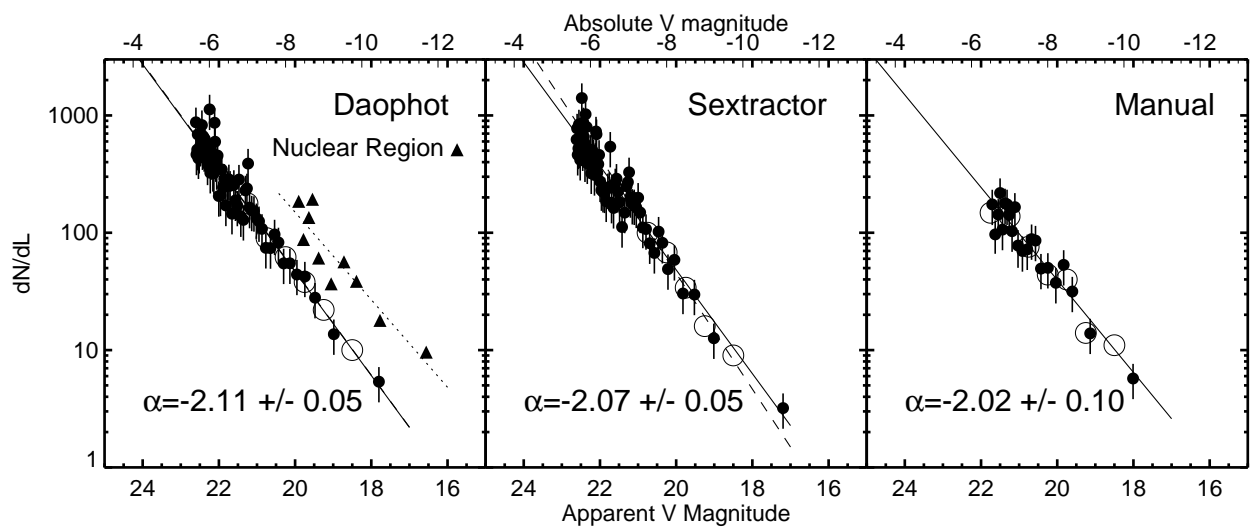}
\caption{Luminosity functions using Daophot, Sextractor, and Manually
selected catalogs for candidate clusters located outside of the nuclear starburst region (circles) and inside the nuclear region (triangles; a constant offset was applied to these data, which are shown for the Daophot catalog only). Variable-binning is used for the solid symbols and fixed-binning for open symbols. The fits are given for the variable-binning data (see text for details). The luminosities have not been corrected for extinction in M83.}
\label{fig:lf_object_selection}
\end{figure}

Figure~\ref{fig:lf_object_selection} shows luminosity functions (hereafter: LF) for clusters outside of the nuclear region, in the F555W filter. The luminosities include our preferred, size-dependent aperture corrections, and have been corrected for extinction in the Milky Way but not for extinction in M83.  Two different binnings are shown in each panel, variable size bins with equal numbers of clusters in each bin (as recommended by Maiz Apellaniz \& Ubeda 2005 and shown with filled circles) and approximately equal size bins with variable numbers of clusters in each bin (open circles).
Variable binning will be our  preferred method throughout this paper, and will be discussed in more detail in Section~4.2. The left panel of Figure~\ref{fig:lf_object_selection} also shows the luminosity function (in F555W) for clusters in the nuclear starburst region (filled triangles).

Each of these luminosity functions can be described, to first order, by a single power law, $\phi(L) \propto L^{\alpha}$.  The best fit value of $\alpha$ is given in each panel for the variable size bins.  These values are the same, within the errors. We find that the exponent $\alpha$ does not change with luminosity, i.e., there is no evidence for a steepening or flattening at brighter magnitudes. We also determine $\alpha$ in several other ways from our cluster catalogs: in different filters, by selecting subsamples which have colors in cluster space and by $U\!-B\!$ color.
These values for $\alpha$ are compiled in Table~\ref{tab:alpha}, and discussed further in \S4.2. We find a power-law index for the luminosity function of clusters in M83 to be $\alpha = -2.04 \pm 0.08$. This is the
mean and standard deviation of all  $\alpha$ values listed in Table~\ref{tab:alpha} (excluding the nuclear dataset at the end of the table). The luminosity function of clusters in the nuclear region have a best fit value of $\alpha=-1.93\pm 0.12$, which is the same as that found for clusters outside of the nuclear region, within the errors.

\subsection{The Impact of Different Assumptions on the Luminosity Function}

Here, we quantify the sensitivity of the luminosity function of clusters in M83 to different selection techniques, assumptions, and corrections that are typically made in the course of the analysis. The following items are arranged in the order in which they affect the power-law index $\alpha$. We note that this order may be different for other datasets with different characteristics, such as lower photometric accuracy or higher extinction than found in our WFC3 M83 data.

\begin{itemize}
\item {\em Red vs.\ Blue Clusters:} The luminosity functions of clusters selected by color, i.e., for red ($U\!-\!B > -0.5$; generally older) and blue ($U\!-\!B < -0.5$; young) has the strongest impact on the power law exponent of the luminosity function in our study, with an RMS~$\approx0.08$. While two of the catalogs (SExtractor and Manual) have values of $\alpha$ for the red and blue samples which are within the uncertainties, the other two (Daophot and Ishape) show differences of $\approx2.5\sigma$, with the blue sample having slightly shallower values for $\alpha$ (see Table~1). This result however, does not account for any extinction in M83. When we correct the luminosity of each cluster by the extinction derived from our dating analysis (described in Section~5.1), the resulting values of $\alpha$ are the same for the red and blue subsamples within the uncertainties.

\item {\em Filter (i.e., Wavelength):} The choice of filter has a relatively minor effect on values of $\alpha$, with an RMS$\approx0.06$ between the four filters of each cluster catalog. While there appears to be a weak trend 
for the shorter wavelengths to have flatter values of $\alpha$, the values for the F336W and F814W filters are within the formal uncertainties.

\item {\em Cluster Selection:} The particular method used to select clusters 
(e.g., manual or automatic, based on different measurements of object size from different software) for this particular data set does not appear to have
a significant impact on the shape of the luminosity function. The average value of $\alpha$ for the four catalogs in the F555W filter, as listed in Table~1, has an RMS $\approx 0.04$. All have the same value of $\alpha$ within the uncertainties.

\item {\em Constant vs.\ Variable Binning:} It has been suggested by Maiz Apellaniz \& Ubeda (2005) that using constant instead of variable size bins 
can cause variations as large as 0.3 in $\alpha$ for small datasets, with values for constant binning being too steep. Here we find a much smaller effect,  with $\alpha$ typically steeper by only $\approx0.03$ for constant binning when compared with variable binning. This indicates that the specific choice of binning has little effect on the shape of our luminosity functions.

\item {\em Mean vs.\ Size-Dependent Aperture Corrections:} While the difference between assuming mean and  size-dependent aperture corrections can result in  large variations in the total magnitude for a given cluster,  the overall affect on $\alpha$ is quite small. We find an RMS $=0.01$ between the values of $\alpha$ when the mean vs.\ size-dependent aperture corrections are used.

\end{itemize}

To summarize, we find that a variety of effects, including color, filter, selection technique, and binning, can all impact the power-law index $\alpha$ of the luminosity function. However, we find that each of these has a small affect, at less than the $\approx0.08$ level. This is almost certainly a lower limit to the uncertainty on $\alpha$ in most extragalactic studies of star forming galaxies, since most observations acquired to date are not as high quality as those presented here, and because the separation of clusters from stars and background galaxies becomes more difficult in more distant galaxies.
We conclude that the luminosity function of star clusters in M83 has relatively
small uncertainties, and can be described by a power law with $\alpha=-2.04\pm0.08$ for $m_V \lea 23$~mag ($M_V \lea -5.5$).

\section{Mass and Age Distributions of Star Clusters in M83}

\subsection{Dating Technique and Results}

The mass and age distributions for a population of star clusters provide the primary window into the formation and disruption of the clusters. We estimate the age, extinction, and mass of each cluster as we have done in previous works (see Fall et~al.\ 2005 and Whitmore et~al.\ 2010 for details),
by performing a least $\chi^2$ fit comparing measurements in five filters (\textit{UBVI},H$\alpha$) with new Charlot \& Bruzual models (CB09) using the WFC3 filter transmission curves with solar and $\approx$2$\times$solar metallicity. We assume a Chabrier (2003) initial stellar mass function (IMF), and a Galactic extinction law (Fitzpatrick 1999). The mass of each cluster is estimated from the observed, extinction corrected $V$ band magnitude and the mass-to-light ratio predicted by the models, assuming a distance modulus
$\Delta(m-M)=28.28$ to M83. If we had adopted the Salpeter (1955) rather than the Chabrier IMF, the $M/L_V$ and hence the masses would increase by
a near constant (age-independent) $\approx40$\%.

\begin{figure}
\epsscale{0.8}
\plotone{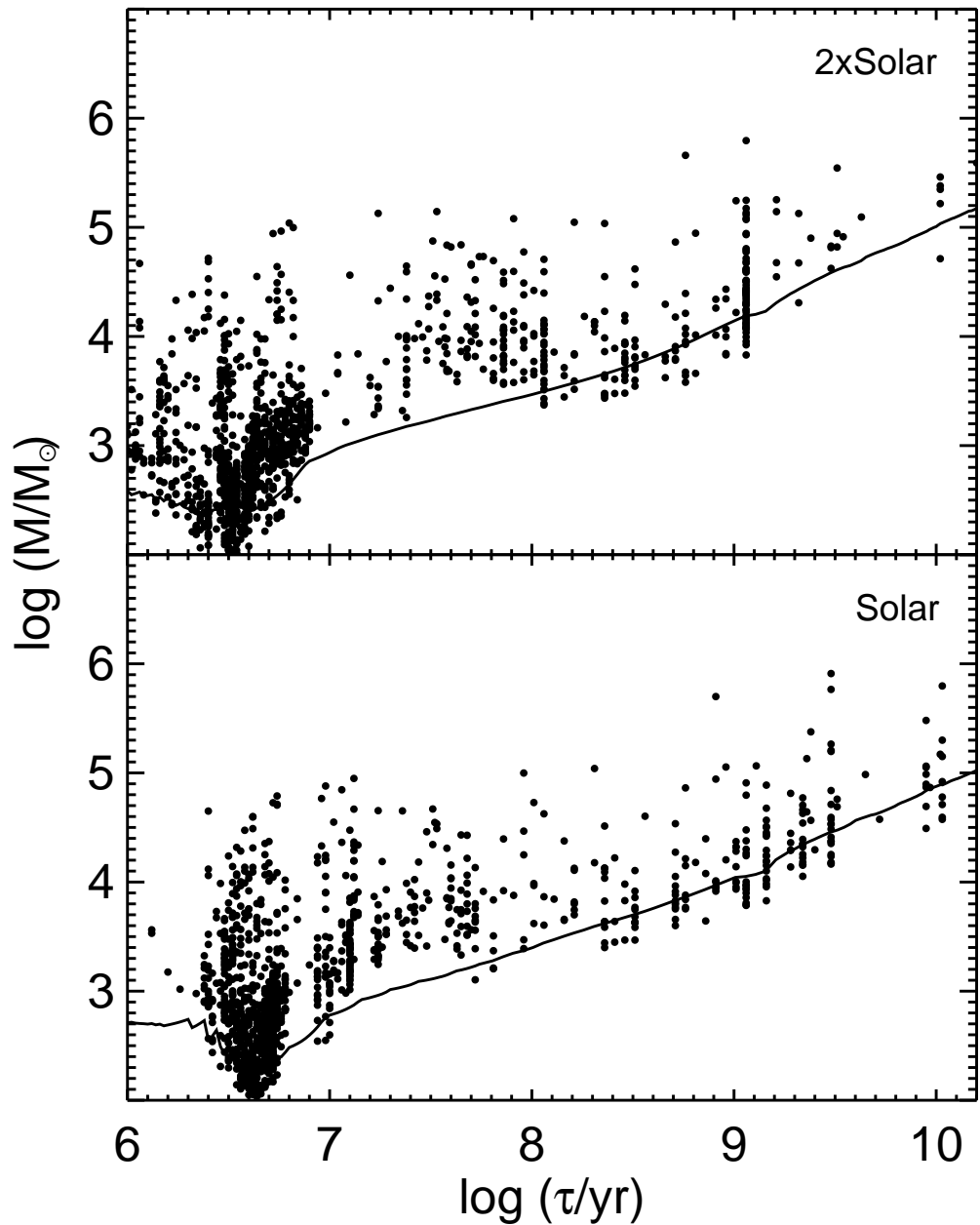}
\caption{Log~$M$ vs.\ $\log\tau$ diagram for clusters in our Daophot selected catalog.  The top panel shows the results when we assume the $Z=0.04$ ($\approx2\times$ solar) metallicity model from Charlot \& Bruzual, and the bottom panel shows the results when the $Z=0.017$ (solar) metallicity model is assumed. The solid line in each panel shows a magnitude limit of $M_V=-6.0$. Note that the broad trends in the distribution of cluster masses and ages are not sensitive to metallicity for ages $\tau \lea 10^9$~yr.}
\label{fig:Mt}
\end{figure}

In Figure~\ref{fig:Mt} we show the mass-age ($M$-$\tau$) diagram of star clusters in M83 that results from the dating analysis of our Daophot catalog,
assuming $2\times$solar (upper panel) and solar metallicity (lower panel).
We find ages for the clusters that span the lifetime of the galaxy. The extinction values in M83 are fairly low, with a typical value of $A_V\approx 0.7$ for $\tau \lea 10^7$~yr clusters located outside the nuclear regions, and $A_V\approx 2$ for clusters inside the nuclear starburst region. The $M$-$\tau$ diagrams show a number of small-scale features, with pile-ups at specific ages and gaps at others. The gap between $7.0 \lea \log(\tau/\mbox{yr}) \lea 7.5$~yr in the upper panel, for example, occurs where the predicted colors loop back on themselves, covering a small region in color
space over a relatively long time, and effectively resulting in a gap. The broad distribution of cluster masses and ages, however, is not greatly affected by these small-scale features, or by the adopted metallicity. We adopt the masses and ages resulting from our comparison with the $2\times$solar metallicity model in the rest of this paper, because this model gives a somewhat better match to our observed colors for the clusters in M83.

Trends in the distribution of cluster masses and ages are apparent from the $M$-$\tau$ diagram. When we look along the horizontal axis, we see that 
the number of clusters is approximately constant in equal bins of $\log\tau$, within a given range of $\log M$, without increasing or decreasing significantly with age. This indicates that the cluster age distribution can be described approximately by a power law, $dN/d\tau \propto \tau^{\gamma}$, with $\gamma\approx-1$. When we look down the vertical axis, we see that 
the number of clusters increases steadily with decreasing mass. This indicates that the cluster mass function can be described approximately by a 
power law, $dN/dM \propto M^{\beta}$, with $\beta$ much steeper than $-1$. We give a more quantitative treatment in the next two subsections.

\subsection{Age Distribution}

\begin{figure}
\plotone{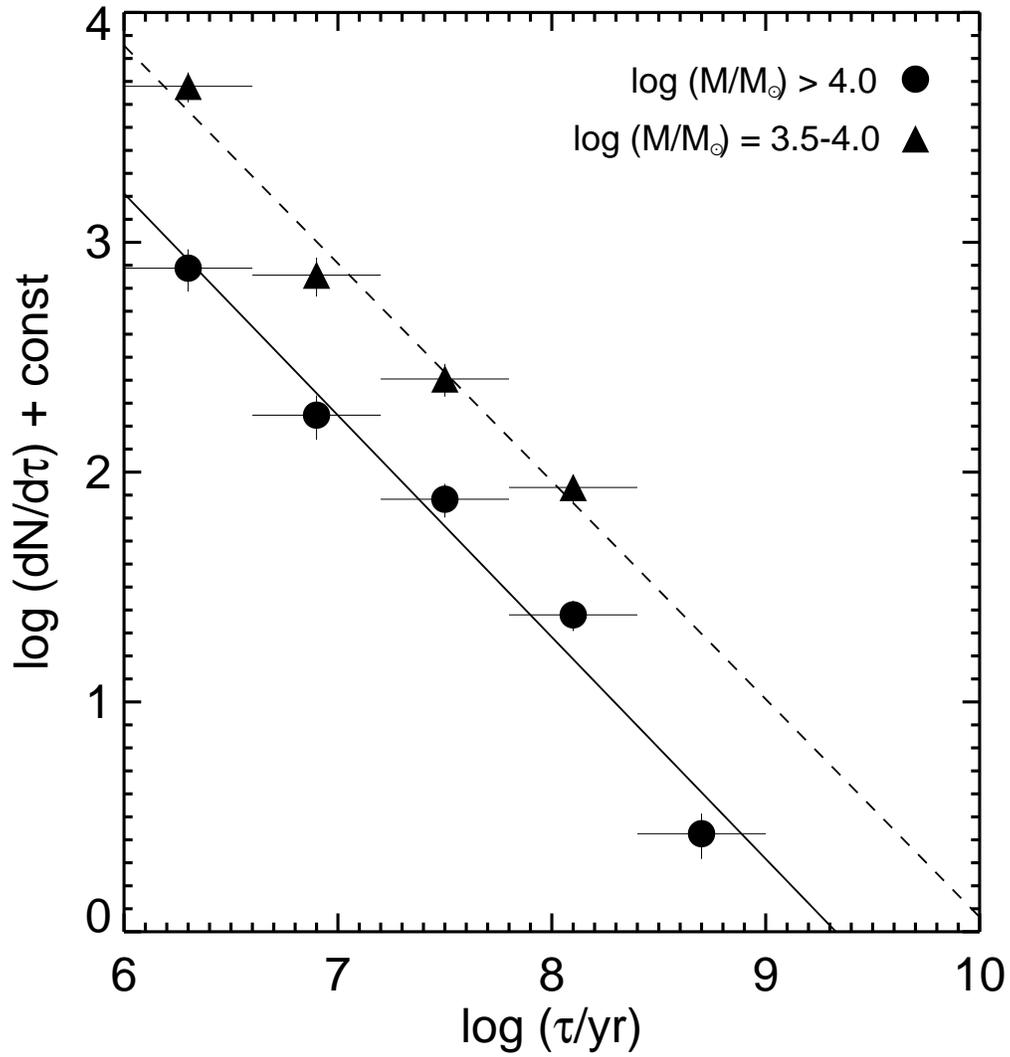}
\caption{The age distribution, $dN/d\tau$ for clusters in M83 in the indicated
intervals of mass.  A power-law index with $\gamma = -0.95$ is shown for both distributions.}
\label{fig:dndt}
\end{figure}

In Figure~\ref{fig:dndt} we show preliminary age distributions for clusters in two different intervals of mass, based on an assumed metallicity of $2\times$solar. A more detailed treatment based on the full set of filters, in particular the addition of F225W, will be included in Chandar \etal\ (in preparation).
These distributions decline steeply, and have nearly identical fits, with $\gamma=-0.95\pm0.10$. If we use the results from the solar metallicity model instead, we find $\gamma=-1.04\pm 0.12$. The $\gamma$ values 
determined from the different catalogs and different mass ranges are all within $\pm0.20$ of those shown in Figure~\ref{fig:dndt}. Based on these experiments, we conclude that the exponent for the age distribution of clusters is approximately $\gamma=-0.9\pm0.2$.

This result, that the age distribution declines steeply for clusters more massive than a $3\times10^3~M_{\odot}$ and younger than a few$\times10^8$~yr, is quite similar to those found recently by Mora et~al.\ (2009), based on a different set of observations (i.e.,  ACS and WFPC2) and a field farther out in M83. Mora et~al.\ found $\approx200$ compact star clusters in two ACS pointings in M83, and estimated ages by comparing \textit{UBVI} measurements with stellar population models (the $U$ band measurements are from WFPC2). They determined the age distribution by counting the number of clusters brighter than a given $V$ band luminosity in different bins of $\log\tau$, and found them to be consistent with a value of $\gamma\approx-0.7$ after converting their luminosity-limited result to a mass-limited one. By contrast, Gieles \& Bastian (2008) claim a flat age distribution with $\gamma\approx0$ from the Mora et~al.\ sample. Their result is based on an indirect technique that plots the mass of the most massive cluster as a function of age in the form $\log M_\mathrm{max}$ vs.\
$\log \tau$, and uses the resulting slope to infer the shape of the age distribution, assuming a constant shape for the mass function over time
(see Chandar et~al.\ 2010a for more details). The Gieles \& Bastian result is dominated by a single data point, the youngest one, where they find a maximum mass of only $M\approx10^3~M_{\odot}$ for clusters younger than $\tau < 10^7$~yr.\footnote{ Figure~7 in Mora et~al.\ (2009) shows masses and ages for clusters in M83, derived by comparing their measured colors with different population synthesis models. In all cases where they derive ages younger than $10^7$~yr, they also find several clusters more massive than $10^3~M_{\odot}$.} However, our observations of M83 clearly reveal a large number of clusters more massive than $M > 10^3~M_{\odot}$
that formed in the last $\tau < 10^7$~yr.

The declining shape, with $\gamma\approx-1$, found for the age distribution 
of young star clusters in M83 is similar to that found in over a dozen other galaxies for $\tau \lea \mbox{few}\times10^8$~yr, including dwarf and massive galaxies, spirals and irregular galaxies, interacting and isolated
galaxies. As summarized in Chandar et~al.\ 2010a, virtually all of the star-forming galaxies studied to date have age distributions with similar declining shapes, with $\gamma$ values between $\approx-0.7$ and
$\approx-1.0$ for mass-limited samples, or have $M$-$\tau$ diagrams that have approximately equal numbers of clusters in equal bins of $\log\tau$, above a given mass. These similarities in over a dozen different galaxies
suggests that it is the disruption rather than the formation of the clusters that
is primarily responsible for shaping the age distribution. It is much more likely 
that clusters in all of these galaxies have had similar disruption histories than it is that they have had similar formation histories, and that we just happen to be observing them all at exactly the time when their formation rates have peaked. Several different processes may disrupt clusters on relatively short timescales $\tau \lea \mbox{few}\times 10^8$~yr, while preserving the shape of the mass function (see Fall et~al.\ 2009; 2010, Chandar et~al.\ 2010b).

\subsection{Mass Function}

\begin{figure}
\epsscale{1.0}
\plotone{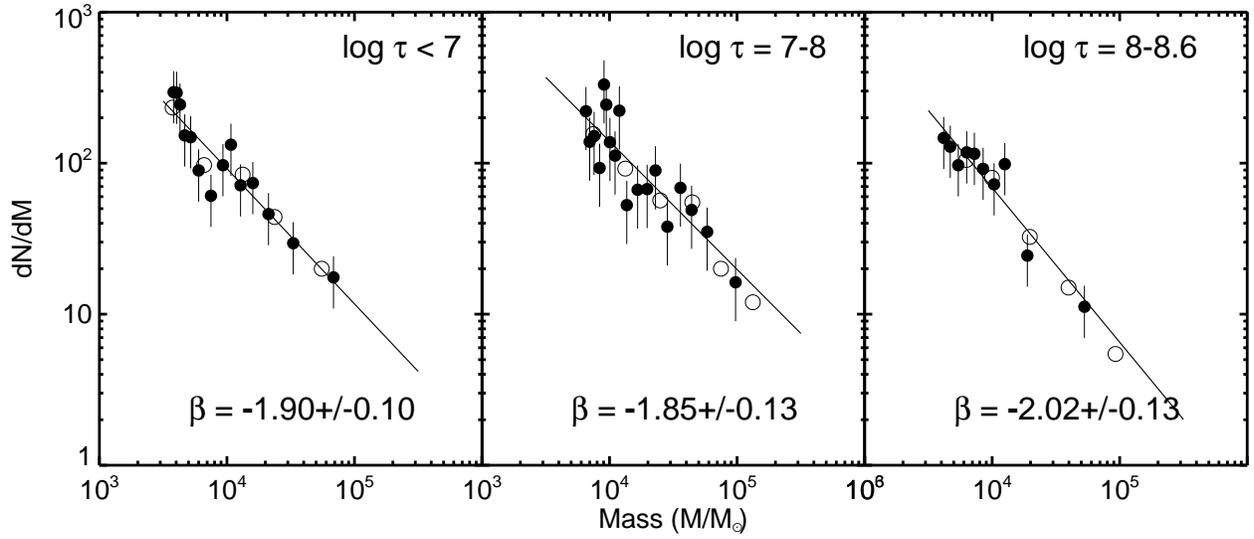}
\caption{Mass function for M83 clusters, based on the Daophot catalog, averaged over the indicated intervals of age. The solid lines show the best fits to the variable-binning data (filled circles), while the open circles show
the fixed-bin data. The given values of $\beta$ are for the variable-binning
data.  See text for more details.}
\label{fig:mf}
\end{figure}

In Figure~\ref{fig:mf}, we present preliminary mass functions for star
clusters in M83 in three different intervals of age: $\log\tau=6$--7, $\log\tau=7$--8, and $\log\tau=8$--8.6, based on our dating analysis. The mass functions, like the luminosity functions, can be described by a power law, $dN/dM \propto M^{\beta}$. The best fit values of $\beta$ for the different
age ranges are given in Figure~\ref{fig:mf}, and are the same within the uncertainties as those found for the luminosity function, with $\beta\approx-2.0$. Like the luminosity functions, the mass functions show no obvious 
deviation from a power law, such as a bend at either the high or low mass end. In a future paper we will present a more detailed treatment based on
the full set of filters, in particular the addition of the F225W filter, and based on different age-dating methods and a wider variety of stellar population models.  We will also include clusters found in a second pointing in M83. We find a power-law index for the mass function of clusters in M83 to be $\beta = -1.94\pm0.16$, for $M\gea \mbox{few}\times10^3~M_{\odot}$ and $\tau \lea 4\times10^8$~yr. This is the mean and standard deviation of all $\beta$ values shown in Figure~\ref{fig:mf}. We find similar results for the three other cluster catalogs described in Section~3.2. The power-law index for the LF ($\alpha$) and for the mass function ($\beta$) are the same within their uncertainties, with $\alpha = -2.04 \pm 0.08$ and  $\beta = -1.94 \pm 0.16$.
While the luminosity and mass functions are sometimes considered to be interchangeable, this is not true for populations of clusters that include a wide range of ages, and hence mass-to-light ratios. In fact, the observed similarity in the values of $\alpha$ and $\beta$, which are both power laws,
is (indirect) evidence that the age and mass distributions are independent of one another, at least for $\tau \lea4\times10^8$~yr (see for e.g., Fall 2006).

\section{Discussion}

In this Section we use the observed shape of the mass function of $\log\tau=8.0$--8.6~yr clusters to investigate two issues related to the formation and evolution of the clusters. First, we assess whether or not there is evidence  for an upper cutoff to the masses with which clusters in M83 can form.
Curvature at the high end of the mass function would provide evidence for such a physical limit. Second, we determine whether there is evidence for the 
early disruption of lower mass clusters when compared with their higher mass counterparts. Curvature or a flattening at the low end of the mass function would provide evidence for the mass-dependent disruption of the
clusters.

\subsection{Is There Evidence for an Upper Mass Cutoff?}

Several recent papers have suggested that there is a cutoff at the high end of the mass function of relatively young ($\tau \lea \mbox{few} \times 10^8$~yr) clusters in some spiral galaxies, such as M51, M83, \& NGC~6946 (e.g., Gieles et~al.\ 2006; Larsen 2009). More recently, Portegies Zwart et~al.\ (2010) suggested that a cutoff $M_C\approx2\times10^5~M_{\odot}$ may be present in {\em all} Milky Way-like spiral galaxies. A cutoff can typically be described by a Schechter function, $\psi(M) \propto M^{\beta} \mbox{exp}(-M/M_C)$ (e.g., Burkert \& Smith 2000; Fall \& Zhang 2001; Jordan et~al.\ 2007), where the number of massive clusters drops exponentially compared with the number of lower mass clusters, i.e., faster than a power law. For example, Schechter functions with $M_C\approx 1-2\times10^6~M_{\odot}$ provide significantly better fits than power laws at the high end of the mass function of old globular clusters (e.g., Burkert \& Smith 2000; Fall \& Zhang 2001; Jordan et~al.\ 2007).

We found in Section~5.3 and in Figure~\ref{fig:mf} that a power law provides a good fit to the mass function of clusters in M83 at different ages.
A K-S test comparing the masses of clusters with $M \geq 8\times 10^4~M_{\odot}$ with a power law  of $\beta=-2.02$ returns a $P$-value of 0.32, i.e., a power law provides an acceptable fit to the upper end of the cluster mass function (note: a $P$-value $< 0.05$ is typically indicative of an
unacceptable fit).

The upper panel of Figure~13 compares the observed mass function with three different values of $M_C$: $10^5~M_{\odot}$, $4\times10^5~M_{\odot}$, and $10^6~M_{\odot}$. This figure suggests that values of $M_C$ that are lower than $\approx10^5~M_{\odot}$ do not provide a good match to the data. This visual impression is confirmed by statistical tests. We are only able to place a lower-limit on $M_C$, with all values $M_C > 1\times10^5~M_{\odot}$ giving acceptable fits, i.e., within a 95\% confidence level, based on formal K-S tests. We conclude that, because a power law provides an acceptable fit, our data do not require a Schechter function to describe them.

\subsection{Is There Evidence for Mass-Dependent Cluster Disruption in M83?}

\begin{figure}
\epsscale{0.8}
\plotone{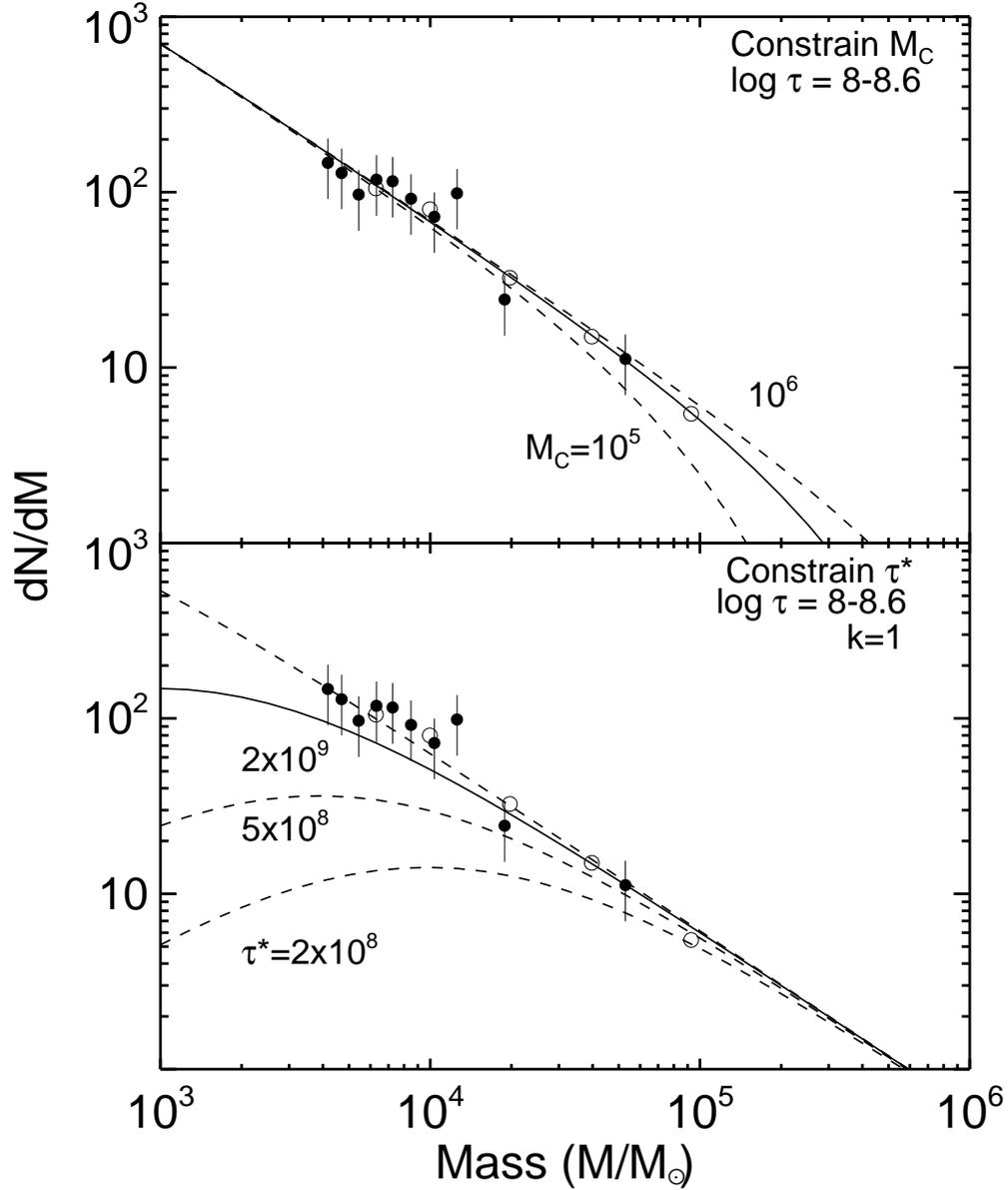}
\caption{The mass function for the $\log\tau=8.0$--8.6~yr clusters in M83 is compared with different predictions for an upper mass cutoff $M_C$ (upper panel), and for different characteristic disruption timescales $\tau_*$ (lower panel). The curves in the upper panel are Schechter functions, $\psi(M) \propto M^{\beta} \mbox{exp}(-M/M_C)$, with $M_C=1\times10^5~M_{\odot}$, $4\times10^5~M_{\odot}$, and $1\times10^6~M_{\odot}$. The curves in the lower panel use equations~(B6)--(B8) from Fall et~al.\ 2009 to predict the evolution of the mass function for a population of $\log\tau=8.0$--8.6~yr clusters, with the indicated disruption timescales $\tau_{*}$, for a linear rate of mass loss ($k=1$).  See text for details.}
\label{fig:mfmodels}
\end{figure}

Next, we use the shape of the mass function to study the dynamical evolution of the clusters. Some processes that disrupt clusters, such as the loss of stars due to (internal) two-body relaxation, will disrupt lower mass clusters earlier than higher mass clusters (at a constant density). Two-body relaxation causes clusters to lose mass at an approximately linear rate (e.g., Fall \& Zhang 2001; see McLauglin \& Fall 2008 for a detailed discussion of
evaporation rates), giving a disruption time that depends on the initial mass $M_0$ of the cluster as $\tau_d(M_0)=\tau_{*}(M_0/M_{*})^{k}$, with $k=1$. Here, $\tau_*$ is the characteristic time it takes to disrupt a $M_*=10^4~M_{\odot}$ cluster. In Figure~\ref{fig:mfmodels} (lower panel) we compare the mass function for log~$\tau=8-8.6$~yr clusters for different values of $\tau_*$: $2\times10^8$~yr, $5\times10^8$~yr, and $2\times10^9$~yr.
Because the mass function follows a power-law without flattening at lower masses, we can only place a lower limit on $\tau_*$, and find $\tau_* \gea 2\times 10^9$~yr. This result does not mean that cluster disruption does not take place until after $2\times 10^9$~yr, only that the disruption of clusters on shorter timescales does not depend strongly on the mass of the clusters.

The results presented here indicate that the cluster age distribution is approximately independent of mass, and that the cluster mass function is approximately independent of age. This means that the bivariate distribution of cluster masses and ages, $g(M,\tau)$, can be written as the product 
of the mass and age distributions: $g(M,\tau) \propto M^{\beta}\tau^{\gamma}$, with $\beta=-1.94\pm0.16$ and $\gamma=-0.9\pm0.2$. This result for M83 is similar to those found for the Magellanic Clouds (e.g., Parmentier \& de Grijs 2008; Chandar et~al.\ 2010b) and for the Antennae (Fall et~al.\ 2009), and indicates that mass-dependent disruption plays
little role, but that mass-{\em in}dependent disruption plays a strong role,  in shaping the masses and ages of clusters in these galaxies for the first $\tau \lea \mbox{few}\times10^8$~yr.

\section{Summary and Conclusions}

We have used observations taken with the newly installed WFC3 camera
on-board the \emph{Hubble Space Telescope} to study star clusters in the nearby spiral galaxy M83. Our main science goals were to determine the luminosity, mass, and age distributions of the clusters, and to use these to understand the formation and disruption of the clusters. In order to accomplish these goals, we used several different methods to select compact star clusters, and assessed the impact that various assumptions and corrections have on the shape of the luminosity function.

We found that the luminosity function of star clusters in M83 can be described by a power law, $\phi(L) \propto L^{\alpha}$, with $\alpha=-2.04\pm0.08$. We found that the selection of the clusters, i.e., the specific criteria that are used to find and catalog them, does not have a significant impact on $\alpha$, with variations on the order of $\approx\pm0.04$.
Other issues that impact $\alpha$, roughly in order of importance are: object color ($\pm0.08$), filter ($\pm0.06$),  binning ($\pm0.03$),  and aperture corrections ($\pm0.01$). 

We estimated the masses and ages of the clusters by comparing measurements of \textit{UBVI},$H\alpha$ magnitudes with new stellar evolution models from Charlot \& Bruzual (2009), and presented preliminary mass and age distributions for the clusters in our field. The age distribution declines steeply, $dN/d\tau \propto \tau^{\gamma}$, with $\gamma=-0.9\pm 0.2$, for clusters more massive than $M \gea \mbox{3}\times10^3~M_{\odot}$ and ages younger than $\tau \lea \mbox{few} \times10^8$~yr, and has the same shape for different masses. These results were interpreted to mean that $\approx80$--90\% of clusters are disrupted every decade in age,
in a manner that does not strongly depend on their mass.

We found that the mass function can be described by a power law, $dN/dM \propto M^{\beta}$, with $\beta=-1.94\pm0.16$, for clusters with $M \gea \mbox{few}\times10^3~M_{\odot}$ and $\tau \lea 4\times10^8$~yr. The observed mass function does not show bends at either the high or low mass end. The mass function does not require an upper mass cutoff, $M_C$, since a power law provides a statistically acceptable fit. We did not find evidence that lower mass clusters are disrupted earlier than higher mass clusters, with a lower limit on any characteristic disruption time of $\tau_* \gea 2\times10^9$~yr for a $10^4~M_{\odot}$ cluster. 

Our results for the luminosity, mass, and age distributions of star clusters in M83 are similar to those found recently for the cluster systems in a growing number of other galaxies (e.g., the Antennae, Fall \etal\/ 2005, 2009, Whitmore \etal\/ 2007; the Large and Small Magellanic Clouds, Chandar \etal\/ 2010b; and several spiral galaxies, Mora \etal\/ 2009). This suggests that the shapes of the mass and age distributions for young cluster systems in nearby star forming galaxies, and hence the formation and disruption mechanisms that control these distributions,  may be relatively ``universal.''

\acknowledgments

We thank Zolt Levay for making the color images used in Figure 1 and Figure 3, and Gustavo Bruzual and Stephane Charlot for providing new cluster SED models for the WFC3 filter set prior to publication.  We also thank the anonymous referee and Mike Fall for helpful comments. RC acknowledges support from NSF through CAREER award 0847467.  This paper is based on Early Release Science observations made by the WFC3 Scientific Oversight Committee.  We are grateful to the Director of the Space Telescope Science Institute for awarding Director's Discretionary time for this program.  Finally, we are deeply indebted to the brave astronauts of STS-125 for rejuvenating \emph{HST}.  This research has made use of the NASA/IPAC Extragalactic Database (NED), which is operated by the Jet Propulsion Laboratory, California Institute of Technology, under contract with NASA.

{\it Facilities:} \facility{HST}.

\clearpage

\clearpage

\clearpage

\clearpage

\clearpage

\clearpage

\begin{deluxetable}{ccccc}
\tablecaption{Fits for Cluster Luminosity Function Exponent $\alpha$\label{tab:alpha}}
\tablewidth{0pt}
\tablehead{
\colhead{Sample} & \colhead{Daophot} & \colhead{Sextractor} & 
\colhead{Ishape} & \colhead{Manual}}
\startdata
all, F555W &$-2.11\pm0.05$ &$-2.07\pm0.13$ &$-2.08\pm0.05$ &$-2.02\pm0.10$\\
$U\!-\!B < -0.5$ (blue) &$-2.01\pm0.07$ &$-2.06\pm0.06$ &$-1.99\pm0.06$ &$-1.94\pm0.12$\\
$U\!-\!B > -0.5$ (red) &$-2.19\pm0.07$ &$-2.08\pm0.06$ &$-2.20\pm0.09$  &$-2.07\pm0.10$\\
cluster-space &$-2.07\pm0.07$ &$-2.03\pm0.05$ &$-2.05\pm0.06$ &$-2.00 \pm0.11$\\
F336W &$-1.96\pm0.07$ &$-2.04\pm0.05$ &$-1.95\pm0.06$ &$-1.84\pm0.07$\\
F438W &$-2.03\pm0.10$ &$-2.05\pm0.11$ &$-2.01\pm0.05$ &$-1.94\pm0.10$\\
F555W &$-2.11\pm0.05$ &$-2.07\pm0.13$ &$-2.08\pm0.05$ &$-2.02\pm0.10$\\
F814W &$-2.04\pm0.10$ &$-2.19\pm0.11$ &$-2.05\pm0.10$ &$-1.96\pm0.10$\\
Average &$-2.06\pm0.07$ &$-2.07\pm0.05$ &$-2.05\pm0.08$ &$-1.97\pm0.07$\\
nuclear, F555W &$-1.93\pm0.12$ &\nodata &\nodata &\nodata\\ 
\enddata
\tablecomments{All values of $\alpha$ are from our preferred variable-binning method, and use total luminosities determined from size-dependent aperture corrections. The formal uncertainties from each fit are given. The mean and standard deviation for all values of $\alpha$, excluding the nuclear region, is $-2.04\pm0.08$. }
\end{deluxetable}

\end{document}